\documentclass[twocolumn]{aastex701}
\usepackage{xspace,multirow,amsmath}

\newcommand{\mrk}{{Mrk~335}\xspace}

\shorttitle{Mrk 335}
\shortauthors{Rogantini, Kara et al.}
\graphicspath{{./}{plot/}}%% This is the end of the preamble.  Indicate the beginning of the
\newcommand{\fuse}{{FUSE}\xspace}
\newcommand{\ginga}{{Ginga}\xspace}

\newcommand{\hst}{{HST}\xspace}

\newcommand{\nicer}{{NICER}\xspace}
\newcommand{\nustar}{{NuSTAR}\xspace}

\newcommand{\swift}{{\it Swift}\xspace}

\newcommand{\xmm}{XMM-{\it Newton}\xspace}
\newcommand{\xrism}{{XRISM}\xspace}

\newcommand{\ergcm}{{\ensuremath{\rm{erg\ cm}^{-2}\ \rm{s}^{-1}\ {\AA}^{-1}}}\xspace}
\newcommand{\ergflux}{{\ensuremath{\rm{erg\ cm}^{-2}\ \rm{s}^{-1}}}\xspace}

\newcommand{\kms}{\ensuremath{\mathrm{km\ s^{-1}}}\xspace}

\newcommand{\NH}{\ensuremath{N_{\mathrm{H}}}\xspace}

\newcommand{\spex}{{\textsc{Spex}}\xspace}
\newcommand{\pion}{{\texttt{pion}}\xspace}

\newcommand{\vout}{\ensuremath{v_{\rm out}}\xspace}

\newcommand{\civ}{\ion{C}{6}\xspace}
\newcommand{\cvi}{\ion{C}{6}\xspace}

\newcommand{\neix}{\ion{Ne}{9}\xspace}
\newcommand{\nex}{\ion{Ne}{10}\xspace}

\newcommand{\ovii}{\ion{O}{7}\xspace}
\newcommand{\oviii}{\ion{O}{8}\xspace}

\newcommand{\sixi}{\ion{Si}{11}\xspace}

\newcommand{\mgix}{\ion{Mg}{9}\xspace}
\newcommand{\mgx}{\ion{Mg}{10}\xspace}
\newcommand{\mgxi}{\ion{Mg}{11}\xspace}

\newcommand{\fexvii}{\ion{Fe}{17}\xspace}

\newcommand{\fexx}{\ion{Fe}{20}\xspace}
\newcommand{\fexxi}{\ion{Fe}{21}\xspace}
\newcommand{\fexxii}{\ion{Fe}{22}\xspace}
\newcommand{\fexxiii}{\ion{Fe}{23}\xspace}

\begin{document}

\title{Variability of the X-ray obscuring wind in Mrk 335 with XMM-Newton/RGS}

\author[orcid=0000-0002-5359-9497]{Daniele Rogantini}
\affiliation{Department of Astronomy and Astrophysics, University of Chicago, Chicago, IL 60637, USA}
\affiliation{MIT Kavli Institute for Astrophysics and Space Research, Massachusetts Institute of Technology, Cambridge, MA 02139, USA}
\email[show]{danieler@uchicago.edu}

\author[orcid=0000-0003-0172-0854]{Erin Kara}
\affiliation{MIT Kavli Institute for Astrophysics and Space Research, Massachusetts Institute of Technology, Cambridge, MA 02139, USA}
\email{ekara@mit.edu}

\author[orcid=0009-0006-4968-7108]{Luigi C.\ Gallo}
\affiliation{Department of Astronomy and Physics, Saint Mary's University, Halifax, NS, Canada}
\email{luigi.gallo@smu.ca}

\author[orcid=0000-0002-9214-4428]{S. Komossa} % never spell out first name
\affiliation{Max-Planck-Institut f\"ur Radioastronomie, Auf dem H\"ugel 69, D-53121 Bonn, Germany}
\email{skomossa@mpifr-bonn.mpg.de}

\author[orcid=0000-0003-4511-8427]{Peter Kosec}
\affiliation{Center for Astrophysics \textbar\ Harvard \& Smithsonian, Cambridge, MA, USA}
\email{peter.kosec@cfa.harvard.edu}

\author[orcid=0009-0001-9034-6261]{Christos Panagiotou}
\affiliation{MIT Kavli Institute for Astrophysics and Space Research, Massachusetts Institute of Technology, Cambridge, MA 02139, USA}
\email{cpanag@mit.edu}

\author[orcid=0000-0002-4794-5998]{Dan Wilkins}
\affiliation{Department of Astronomy, The Ohio State University, Columbus, OH, USA}
\email{wilkins.401@osu.edu}

\author[orcid=0000-0001-9735-4873]{Ehud Behar}
\affiliation{Department of Physics, Technion, Haifa 32000, Israel}
\email{behar@physics.technion.ac.il}

\author[orcid=0000-0002-0568-6000]{Joheen Chakraborty}
\affiliation{MIT Kavli Institute for Astrophysics and Space Research, Massachusetts Institute of Technology, Cambridge, MA 02139, USA}
\email{joheen@mit.edu}

\author[orcid=0000-0002-9961-3661]{Dirk Grupe}
\affiliation{Department of Physics, Geology, \& Engineering Technology, Northern Kentucky University, Highland Heights, KY, USA}
\email{gruped1@nku.edu}

\author[orcid=0000-0002-4992-4664]{Missagh Mehdipour}
\affiliation{Department of Astronomy, University of Michigan, Ann Arbor, MI, USA}
\email{missagh@umich.edu}

\author[orcid=0000-0003-2532-7379]{Ciro Pinto}
\affiliation{INAF -- IASF Palermo, Via Ugo La Malfa 153, I-90146 Palermo, Italy}
\email{ciro.pinto@inaf.it}

\author[orcid=0000-0001-7630-8085]{Irina Zhuravleva}
\affiliation{Department of Astronomy and Astrophysics, University of Chicago, Chicago, IL 60637, USA}
\email{zhuravleva@uchicago.edu}

\begin{abstract}
Transient X-ray obscuration in Seyfert 1 galaxies likely arises from clumpy accretion-disk winds near the broad-line region (BLR), but the wind structure and short-timescale variability are difficult to measure because high-resolution spectra are often suppressed during deep low states. We analyse a coordinated \xmm/\nustar\ campaign on \mrk\ in June 2021, with long-term \swift\ monitoring, capturing the source in an intermediate-flux state with strong RGS absorption features. We model the broadband SED to determine the ionising continuum for self-consistent photoionisation modelling of the RGS spectra. The stacked RGS spectrum requires three photoionised absorbers with $\log\xi \simeq 3.69$, 2.97, and 1.91, outflowing at $|v_{\rm out}|\simeq 5800$, 3200, and 2100 km s$^{-1}$, respectively. Their properties are consistent with the three-phase obscurer reported in 2009, indicating that a similar multi-phase obscuring wind can persist over decade timescales. Using five consecutive RGS observations, we track the wind evolution on day timescales and find strong variability in column density and ionisation in all phases, together with smaller but coherent velocity changes. During a flare, the low-ionisation phase shows a significant drop in opacity, while in the subsequent epoch all phases show increased outflow velocities, suggesting a possible connection between continuum variability and changes in the line-of-sight absorber. The high-ionisation phase responds most directly to changes in ionising luminosity, while the lowest-ionisation phase shows at most a delayed response. Order-of-magnitude constraints place the obscurer at BLR scales, $\sim10^{3}$--$10^{5},R_{\rm g}$, with kinetic power potentially reaching the percent level of $L_{\rm bol}$ for plausible assumptions on geometry and clumpiness.
\end{abstract}



\section{Introduction}
\label{sec:introduction}

Active galactic nuclei (AGN) are among the most variable persistent sources in the Universe, exhibiting dramatic flux changes across a broad range of timescales and wavelengths \citep{Kara25}. The most pronounced variability often occurs in the X-ray band, where the flux can vary by nearly two orders of magnitude within just a few weeks \citep{Ponti12b,Komossa26}. This variability may be driven by transient absorption events caused by clouds transiting the line of sight \citep{Kaastra14,Komossa20,Gallo21,Kara21}, or by intrinsic changes in the geometry or activity of the X-ray-emitting corona \citep{Kara16,Wilkins22}. The competition between intrinsic variability and variable absorption presents a major observational challenge in interpreting AGN light curves and spectra, as these processes can operate simultaneously or mimic one another.

High-resolution X-ray spectroscopy is essential to disentangle absorption from intrinsic spectral changes, particularly through the detection of narrow absorption lines produced by ionised outflows. Narrow-line Seyfert~1 (NLS1) galaxies are especially valuable in this context due to their rapid and large-amplitude X-ray variability \citep{Gallo18}. More than half of Seyfert~1 galaxies exhibit rich soft X-ray spectra marked by absorption lines from warm absorbers, which span a wide range of ionisation parameters, form multiphase structures, and show outflow velocities of a few hundred up to a few thousand km/s \citep{Blustin05, Laha21, Gallo23}. In addition, a handful of Seyfert~1 galaxies reveal absorption features from low-ionisation ultra-fast outflows (UFOs) \citep{Longinotti15,Xu23,Xu24}. These winds are of central interest because they may carry significant mass, momentum, and energy, thereby contributing to AGN feedback on their host galaxies.

 In parallel, multiwavelength campaigns, most notably the \textit{AGN STORM} series \citep{deRosa15,Kara21}, have revealed episodes of strong, transient obscuration (``obscurers'') that can heavily suppress the soft X-ray emission and may represent a distinct form of mass ejection in AGN. These phenomena were established in detail in sources such as Mrk~335 and NGC~5548, where coordinated \xmm\ and \hst\ observations uncovered a newly emerging, partially covering outflow that strongly absorbed the soft X-ray continuum while imprinting deep, blueshifted UV absorption lines \citep{Longinotti13, Kaastra14}. Another well-known case is the NLS1 WPVS~007: as its X-ray flux became very weak, broad UV absorption emerged in \fuse spectra \citep{Leighly09,Leighly15}. Such obscuring winds are markedly different from the commonly observed warm absorbers: they display higher column densities and velocities of several thousand km\,s$^{-1}$, and are launched from the inner broad-line region (BLR) rather than from the narrow-line region (NLR) \citep{Kriss19}. The duration of these obscuration events varies widely, from days to months (e.g.\ Mrk~766, NGC~3227) to several years, as in the decade-long episode in NGC~5548 \citep{Mehdipour22}. An increasing number of similar events have been reported in Seyfert~1 galaxies \citep[e.g.][]{Markowitz14,Ebrero16b,Serafinelli21,Mao22a,Mao22b,Wang22,GraftonWaters23}. A particularly well-observed example is NGC~6814, where a long \xmm\ exposure captured a full obscurer eclipse lasting about 70~ks \citep{Gallo21,Gonzalez24,Pottie23,PothierBogoslowski25}.

Key insights into obscuring winds were recently obtained by the \textit{AGN STORM~2} campaign on Mrk~817, which caught the source during an unexpected obscuration phase \citep{Kara21}. The obscuration was consistent with a dust-free, ionised, partially covering absorber located at the inner BLR \citep[see also][]{Miller21}. Four \xmm\ observations were taken simultaneously with high-cadence \hst\ monitoring; one of them coincided with a brief flux increase in April~2021, revealing a rich spectrum of narrow absorption lines. \cite{Zaidouni24} detected a three-component photoionised wind with velocities of 4000--6000~km/s, consistent with UV absorption lines at the same velocities, confirming a common X-ray/UV origin. Subsequent \nicer\ monitoring attributed the observed variability to combined changes in the obscurer column density (by more than an order of magnitude) and intrinsic continuum flux, indicating a clumpy disk wind whose transparency and ionisation evolve with luminosity \citep{Partington23}.

Despite extensive observational efforts over the past decade, the physical nature of these obscurers remains poorly understood. They may represent discrete clumps lifted from the accretion disk by magnetic or radiative forces, or denser phases embedded within a continuous disk wind. Their connection to other classes of AGN winds, such as warm absorbers or ultra-fast outflows, is still unclear. The multiphase structure, dynamics, and global covering fraction of the obscuring gas are also uncertain. In addition, their optically thick nature often suppresses the soft X-ray band, limiting detailed spectral diagnostics with current observatories. Only in a few cases, such as Mrk~817 during temporary flux increases, has it been possible to probe the ionisation structure of the wind. However, these opportunities are sporadic and short-lived, preventing a systematic characterisation of obscurer properties and variability.

\begin{table*}[t]
\centering
\caption{\xmm, \nustar, and \swift observations of Mrk\,335 in June 2021 including their gross exposure before filtering. We list only the \swift\ observations used for the SED modelling.}
\label{tab:log}
\begin{tabular}{c c c c c}
\hline\hline
Epoch & Mission & ObsID & Start Date (UTC) & Exposure (ks) \\
\hline
\multicolumn{5}{c}{\textbf{XMM-Newton + NuSTAR}} \\
\hline
\multirow{2}{*}{1} & XMM-Newton & 0842761401 & 2021-06-14 06:03:03 & 93.9 \\
                   & NuSTAR     & 60502006002 & 2021-06-14 03:51:07 & 51.4 \\
\hline
\multirow{2}{*}{2} & XMM-Newton & 0842760201 & 2021-06-16 05:52:43 & 93.9 \\
                   & NuSTAR     & 60502006004 & 2021-06-16 05:46:11 & 47.5 \\
\hline
\multirow{2}{*}{3} & XMM-Newton & 0842761101 & 2021-06-18 05:40:24 & 94.0 \\
                   & NuSTAR     & 60502006006 & 2021-06-18 06:01:09 & 47.3 \\
\hline
\multirow{2}{*}{4} & XMM-Newton & 0842761201 & 2021-06-20 05:34:44 & 93.9 \\
                   & NuSTAR     & 60502006008 & 2021-06-20 06:21:10 & 49.4 \\
\hline
\multirow{2}{*}{5} & XMM-Newton & 0842761301 & 2021-06-22 05:42:46 & 93.0 \\
                   & NuSTAR     & 60502006010 & 2021-06-22 05:06:09 & 43.1 \\
\hline
\multicolumn{5}{c}{\textbf{Swift Monitoring}} \\
\hline
-- & Swift & 00013544097 & 2021-06-13 02:28:35 & 1.68 \\
-- & Swift & 00013544098 & 2021-06-14 03:53:35 & 0.89 \\
-- & Swift & 00013544099 & 2021-06-15 03:58:36 & 0.85 \\
-- & Swift & 00013544100 & 2021-06-16 13:18:36 & 0.95 \\
-- & Swift & 00013544101 & 2021-06-17 14:43:36 & 0.86 \\
-- & Swift & 00013544103 & 2021-06-18 08:19:35 & 1.03 \\
-- & Swift & 00013544102 & 2021-06-19 00:09:35 & 0.99 \\
-- & Swift & 00013544104 & 2021-06-20 16:09:50 & 0.44 \\
-- & Swift & 00013544105 & 2021-06-21 22:09:36 & 1.00 \\
-- & Swift & 00013544106 & 2021-06-23 01:31:35 & 0.47 \\
\hline
\end{tabular}
\end{table*}

Our recent joint \xmm/\nustar\ campaign of \mrk\ opens a new window to study obscuration in unprecedented detail. \mrk\ ($z = 0.0258$, $M_{\rm BH} = 2.7 \times 10^7~M_\odot$; \citealt{Grier12}) exhibits extreme X-ray variability driven by both absorption and changes in coronal geometry, making it a benchmark Seyfert~1 galaxy for studying accretion and outflow processes. Once among the brightest X-ray AGN \citep{Tananbaum78,Bianchi01}, \mrk\ experienced a dramatic drop in flux in 2007 \citep{Grupe08}, followed by a long transition phase and then a prolonged low-flux state punctuated by persistent flickering and episodic high-amplitude flares, sometimes brightening by a factor of $\sim$50 on week-long timescales \citep{Grupe12,Wilkins15b,Gallo19,Komossa20,Tripathi20}. The source typically alternates between three flux regimes: a high state, in which the spectrum appears nearly unabsorbed as before 2007; a low state, in which the soft X-ray emission is almost completely suppressed; and an intermediate (``mid'') state, characterised by strong absorption lines from an obscuring wind \citep{Longinotti13,Liu21}. In June~2021, five consecutive \xmm\ observations captured \mrk\ in this intermediate state, providing an exceptional opportunity to resolve its rich absorption-line spectrum with the Reflection Grating Spectrometer (RGS) and to investigate in detail the variability and structure of its obscuring wind.

In Section~\ref{sec:observation}, we describe the observations and data reduction. The analysis of the broadband spectral energy distribution (SED) is presented in Section~\ref{sec:sed}, while Section~\ref{sec:rgs} focuses on the high-resolution RGS spectra and the signatures of ionised outflows. The results on the variability and properties of the obscuring wind are discussed in Section~\ref{sec:discussion} and summarised in Section~\ref{sec:conclusion}. The spectral analysis was performed using the \textsc{SPEX} package (v3.08.01; \citealt{Kaastra96,Kaastra20}), using a hybrid statistical approach: the $C$-statistic for X-ray fitting and the standard $\chi^2$ statistic for optical/UV data. Unless stated otherwise, all errors are quoted at the 1$\sigma$ confidence level. Spectra were binned following the optimal binning algorithm of \cite{Kaastra16} to preserve spectral resolution while ensuring statistically robust fits.

\section{Observations and data reduction}
\label{sec:observation}

\mrk\ was observed in June 2021 as part of a coordinated campaign consisting of five \xmm\ observations obtained simultaneously with \nustar\ (PI: E.~Kara; \xmm\ proposal ObsID: 08427601). The observations span a total of eight days, providing broadband X-ray coverage across multiple epochs. The observation log is given in Table~\ref{tab:log}. This core program was complemented by an extensive \swift\ monitoring campaign (Swift Key Program, Cycle~15, PI: E.~Kara; proposal number: 1518037), which places the joint \xmm/\nustar\ exposures in a broader temporal context. The observational setup of each instrument, together with the details of the data reduction and processing, is described below.

\subsection{XMM-{\it Newton}} \label{sec:xmm}
The campaign comprised five \xmm\ observations performed over eight days between 2021-06-14 and 2021-06-22, each with a nominal exposure of $\sim94$\,ks.\footnote{Dates refer to the observation start times listed in Table~\ref{tab:log}.} We reduced the data using SAS v21.0 and the latest calibration files, including the energy-dependent correction described in \citet{Furst22}. 

\begin{figure}[t!]
\centering
\includegraphics[width=\hsize]{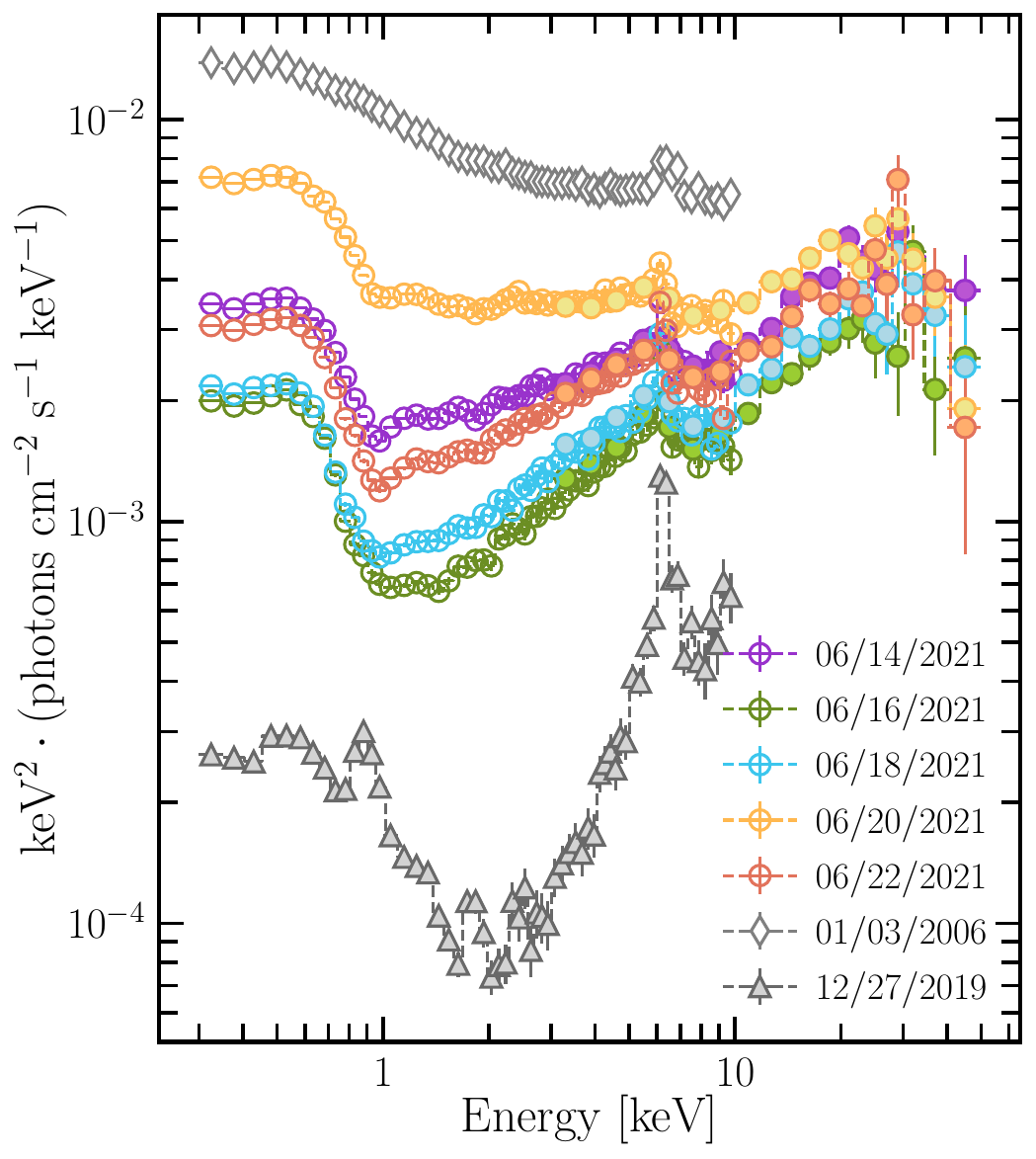}
\caption{
EPIC-pn and \nustar\ spectra of \mrk\ obtained during the 2021 campaign.
Open symbols show the EPIC-pn data ($0.3$--$10$~keV), while filled symbols indicate the simultaneous \nustar\ observations ($3$--$50$~keV).
The coloured datasets correspond to the five 2021 campaign epochs, using the same colour scheme adopted for the light curves in Fig.~\ref{fig:pn_nu_lc}.
For comparison, we also show one of the lowest-flux EPIC-pn spectra ever observed (2019) and one of the highest-flux spectra (2006), just before the onset of the obscuration phase in \mrk.
}
\label{fig:pn_nu}
\end{figure}

\begin{figure*}[ht!]
\centering
\includegraphics[width=\hsize]{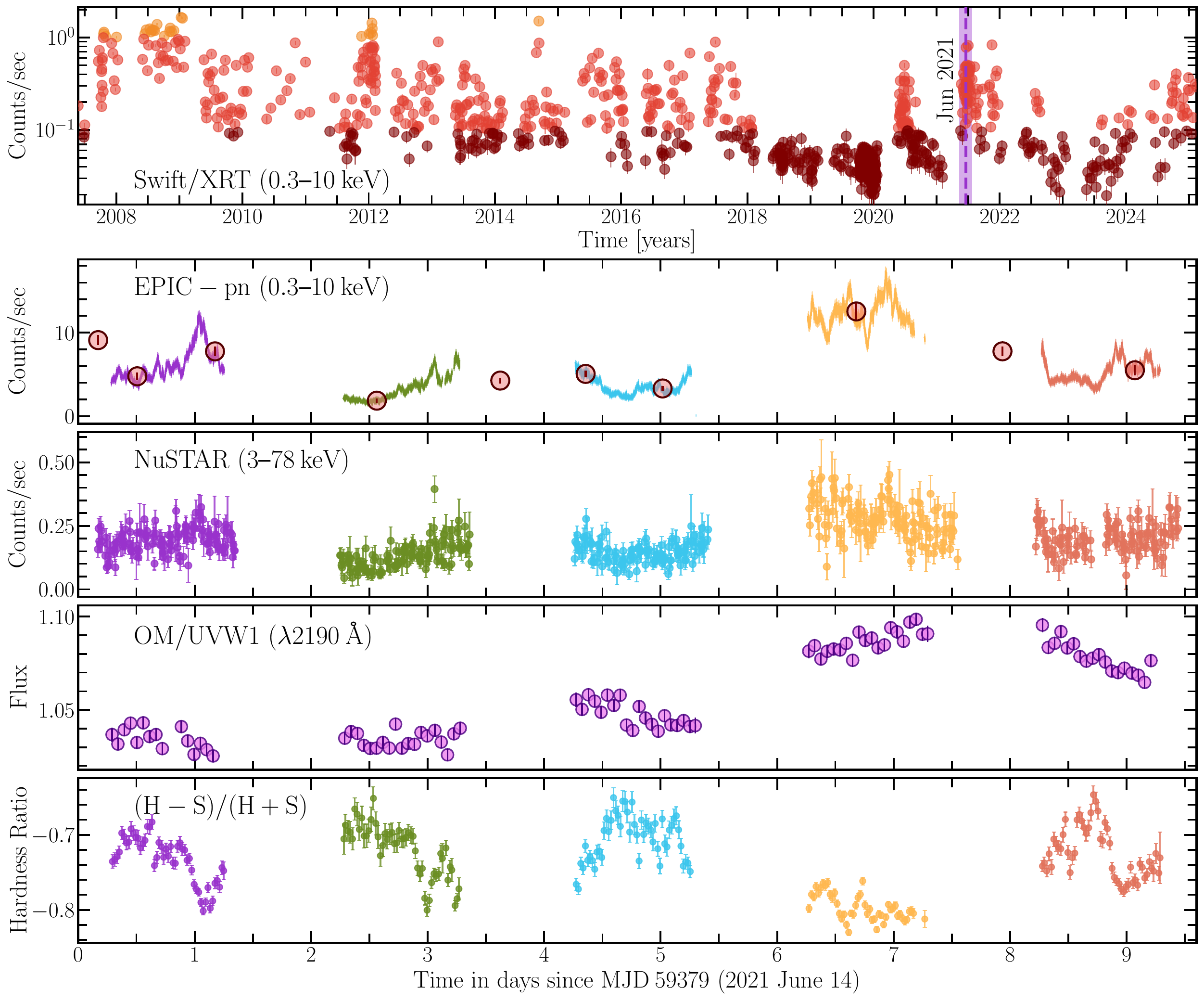}
\caption{
Multiwavelength light curves of \mrk.
\textit{Top panel:} Long-term \swift/XRT monitoring up to May 2025. The vertical dashed line marks the epoch of the joint \xmm\ and \nustar\ campaign. We indicate the approximate high-, mid- and low- flux states with the color palette yellow, red and brown.
\textit{Second panel:} \xmm/EPIC-pn light curves in the $0.3$--$10\ \mathrm{keV}$ band for the five epochs observed in 2021. The \swift/XRT points (shown as dark, lightly filled circles) are overplotted after rescaling with \texttt{WebPIMMS}\footnote{\url{https://heasarc.gsfc.nasa.gov/cgi-bin/Tools/w3pimms/w3pimms.pl}} to match the EPIC-pn count rate.
\textit{Third panel:} \nustar\ FPMA light curve in the $3$--$78\ \mathrm{keV}$ band.
\textit{Fourth panel:} Optical/UV light curve from the OM UVW1 filter, shown in units of $10^{-14}\ \mathrm{erg\ cm^{-2}\ s^{-1}\ \AA^{-1}}$.
\textit{Bottom panel:} Hardness ratio defined as $HR=(H-S)/(H+S)$, where $S$ and $H$ are the \xmm/EPIC-pn count rates in the 0.3--2~keV and 2--10~keV bands, respectively.
}
\label{fig:pn_nu_lc}
\end{figure*}

EPIC-pn (European Photon Imaging Camera; \citealt{Struder01}) was operated in Large Window mode. Periods of high background were excluded by rejecting intervals with count rate $> 0.5$\:counts\:s$^{-1}$ in a high-energy band (10–12 keV) using the {\tt \#XMMEA\_EP} filter. A high-energy single-event light curve (\texttt{PATTERN = 0}, PI between 10–12 keV) was extracted over the full chip to identify flares. Source spectra were extracted from a circular region of radius $30''$ centred on the target. The background was extracted from a same-size circular region on the same CCD, selected from a nearby source-free area and avoiding chip edges where the instrumental Cu~K$\alpha$ line is enhanced. Response matrices and ancillary response files were generated with \texttt{rmfgen} and \texttt{arfgen}, using the option \texttt{applyabsfluxcorr=yes} to apply the calibration correction tied to \nustar\ (see \citealt{Furst22}). After filtering, the typical clean EPIC-pn exposure per observation is $\sim70$\,ks ($\approx75$\% of the nominal). The resulting EPIC-pn spectra from June 2021 are shown in Fig.~\ref{fig:pn_nu}, while the light curves and hardness ratios are presented in Fig.~\ref{fig:pn_nu_lc}. The EPIC-pn observations taken in Large Window mode are affected by pileup. We checked the effect using the task \texttt{epaplot}. For each observation, we extract the source spectra from an annulus with the same outer radius $30''$ and an inner radius between $7.5''$ and $10''$ depending on the count rate and pileup intensity. Comparing the annular (pile-up–mitigated) spectra with the standard circular extractions, we find that pile-up affects the spectral shape only below 1~keV; above 1~keV the spectra are consistent within uncertainties for all epochs. Since we only use the EPIC pn in the energy band between 1.5 keV and 10 keV we use the spectra extracted by the circular region which guarantee the identical spectral fit and provide a higher signal to noise.  

%-----------------------------------------------------------------
%									Single Figure of PN spectra
%-----------------------------------------------------------------

The RGS (Reflection Grating Spectrometers; \citealt{denHerder01}) data were processed with the standard {\tt rgsproc} pipeline. We excluded time intervals when the background count rate in CCD~9 exceeded 0.2~counts\,s$^{-1}$. In the spectral fits, we fitted the RGS1 and RGS2 spectra simultaneously over the 7–37~\AA\ band.

The Optical Monitor (OM; \citealt{Mason01}) was operated in image/fast mode. Throughout, only the UVW1 filter (212~nm) was used, with exposures of 4.4~ks each. Photometry was performed with {\tt omichain} using a circular aperture of diameter $12''$ and standard settings.

\subsection{\nustar} \label{sec:nustar}
\mrk\ was observed simultaneously with \nustar\ \citep{Harrison13}, with exposures ranging from 43 to 51\,ks. The data were reduced using the \nustar\ Data Analysis Software ({\tt nustardas} v1.9.2) and the latest calibration files. Event lists from the two focal plane modules (FPMA and FPMB) were cleaned and calibrated with {\tt nupipeline}, applying standard screening criteria. Periods of elevated background during passages through the South Atlantic Anomaly (SAA) were excluded using the default SAA filtering mode, and the depth correction was applied to reduce the internal high-energy background.

Source spectra were extracted from circular regions of radius $60''$ centred on the target, while background spectra were taken from nearby source-free regions of equal size on the same detector. The {\tt nuproducts} task was used to generate the source and background spectra, together with the corresponding response files. In the spectral analysis, we used the 3--50\,keV band and fitted FPMA and FPMB simultaneously, allowing for a cross-normalisation constant between the two modules. The resulting \nustar\ spectra are shown in Fig.~\ref{fig:pn_nu}, while the light curves are presented in Fig.~\ref{fig:pn_nu_lc}.

\subsection{\swift} \label{sec:swift}
During the June 2021 \xmm\ campaign, \swift\ \citep{Gehrels04} monitored \mrk\ with a cadence of one to two short snapshots per day, with occasional gaps due to gamma-ray burst triggers. The observations were obtained as part of a Cycle~15 Key Project, which started observing \mrk\ on 2020 June 8. Individual visits typically had exposures of 0.5--1\,ks.

The \swift\ X-ray Telescope \citep[XRT;][]{Burrows05} was operated in photon counting mode. Light curves were generated using the online XRT data products tool \citep{Evans07,Evans09}. Standard processing was performed with {\tt xrtpipeline}, which corrects for bad pixels, vignetting, and point-spread function effects to produce cleaned event files. XRT light curves in different energy bands were then extracted for each snapshot. The resulting \swift\ XRT light curve is presented in Fig.~\ref{fig:pn_nu_lc}.

The \swift\ UV/Optical Telescope \citep[UVOT;][]{Roming05} observed \mrk\ with its six primary filters ($V$, $B$, $U$, $UVW1$, $UVM2$, $UVW2$) in end-weighted filter mode (\texttt{0x224c}), using a weighting scheme of 4:3:2:1:1:1 (from $UVW2$ through $V$). Aperture photometry was performed with {\tt uvotsource} using a circular aperture of radius $5''$, applying standard instrumental corrections and calibrations \citep{Poole08}. For broadband spectral fitting with \spex, filter count rates and the corresponding response matrices were generated for each filter.

\section{Spectral Energy Distribution} \label{sec:sed}
%-----------------------------------------------------------------
%                              SED
%-----------------------------------------------------------------
\begin{figure*}[ht!]
\centering
\includegraphics[width=\hsize]{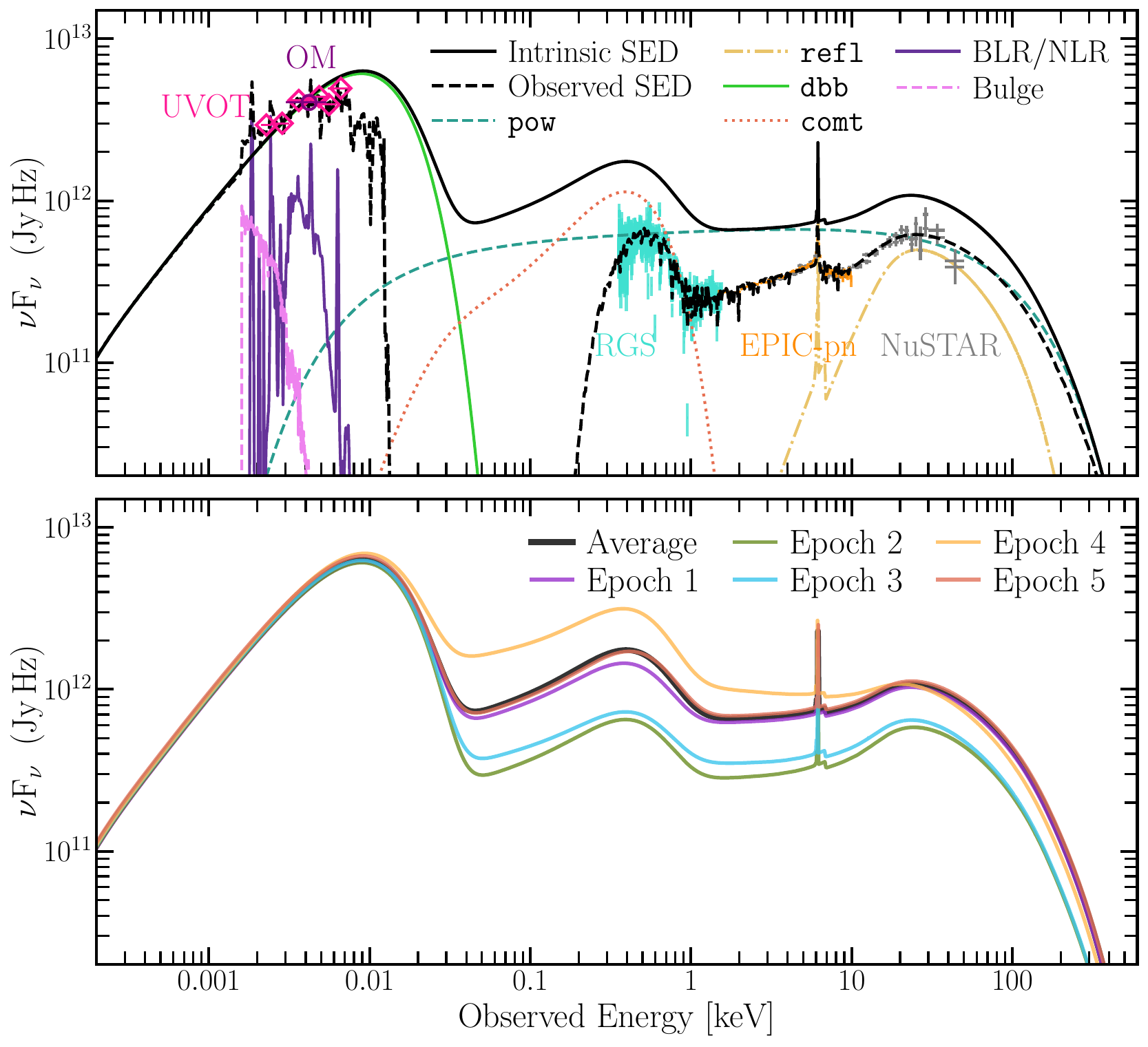}
\caption{Spectral energy distribution (SED) modelling of \mrk.
\textit{Top:} Time-averaged SED modeling with the best-fit model components overplotted (see text for a description of each component), together with the multiwavelength data used in the modelling: XMM/EPIC-pn (orange), \nustar\ (gray), \swift/UVOT (magenta), XMM/OM (purple), and XMM/RGS (cyan). The intrinsic SED (all absorption removed) is shown as a solid black line, while the absorbed (observed) SED is shown as a dashed black line. For visual clarity, the model curves are displayed at an arbitrary energy resolution.
\textit{Bottom:} Best-fit SED models for the individual XMM/\nustar\ epochs in June~2021.
}
\label{fig:sed}
\end{figure*}

In this section we present our modelling of the SED of \mrk, considering both the individual epochs and the time-averaged SED. To derive the intrinsic AGN continuum from the optical/UV to the X-ray band, we simultaneously account for the host-galaxy starlight, UV/optical reddening, and line-of-sight X-ray absorption toward the nucleus. The physical components included in the fits (accretion disc, Comptonisation, reflection, and host-galaxy contributions) are summarised in Table~\ref{tab:sed} together with their best-fit parameter values. The cosmological redshift (\texttt{reds}) was fixed at $z=0.0258$ \citep{Grier12}, corresponding to a luminosity distance of $D_{\rm L}=112.7$~Mpc (for $H_{0}=70~\rm km\,s^{-1}\,Mpc^{-1}$, $\Omega_{\Lambda}=0.70$, and $\Omega_{\rm m}=0.30$). The abundances of all absorption components were set to the protosolar values of \citet{Lodders09}. The resulting average intrinsic SED is shown in Fig.~\ref{fig:sed}, while the bottom panel displays the best-fit SED for each epoch in June~2021. Below we describe the individual components adopted in the SED modelling.

\subsection{Primary optical-UV-X-ray emission of the AGN}
\label{sec:sed_primary}

We modelled the observed primary hard X-ray continuum with a power-law component (\texttt{pow}), which mimics the Compton up-scattering of disc photons in an optically thin, hot corona \citep{Wilkins15a,Keek16}. Based on the analysis of the \nustar\ observations, a high-energy exponential cut-off was applied at $150$\,keV, consistent with the value reported by \citet{Keek16}. In addition, a low-energy cut-off was imposed to avoid exceeding the energy of the seed disc photons, by coupling the turnover energy to the temperature of the disc blackbody component. The photon index $\Gamma$ varies across the observations, ranging from 1.94 to 2.08.

In addition to the primary power-law continuum, the spectrum below 2~keV exhibits a clear ``soft excess'' \citep{Pounds87,ONeill07}. Its physical origin remains debated and is often attributed to warm Comptonisation of disc photons in a ``warm corona'' \citep[see also][]{Magdziarz95,Done12,Petrucci18} and/or relativistically blurred ionised reflection from the accretion disc \citep[e.g.][]{Crummy06,Garcia13a}. Here we model this feature with a phenomenological Comptonisation component (\texttt{comt} in \spex; \citealt{Titarchuk94}), without attempting to distinguish between physical scenarios. This approach has been successfully applied in similar contexts \citep[e.g.\ NGC~3227;][]{Mehdipour21} and is sufficient to capture the ionising continuum required for our photoionisation modelling of the outflows, while being computationally more efficient.

The free parameters of \texttt{comt} are its normalisation, the seed-photon temperature ($T_{\rm seed}$), the electron temperature ($T_{\rm e}$), and the optical depth ($\tau$) of the scattering plasma. In our modelling, $T_{\rm seed}$ was tied to the maximum temperature of the accretion disc ($T_{\rm max}$) described by the \texttt{dbb} model, which assumes a geometrically thin, optically thick Shakura--Sunyaev accretion disc \citep{Shakura-Sunyaev73}. This disc component accounts for the optical/UV emission of \mrk. Because the disc temperature is not expected to vary significantly on week-long timescales (see also the OM light curve in Fig.~\ref{fig:pn_nu_lc}), and given its strong degeneracy with the \texttt{dbb} normalisation due to nearby continuum absorption by neutral hydrogen, we determined $T_{\rm dbb}$ from the average spectrum and restricted it in the single-epoch fits to lie within the uncertainty of that measurement. We found a best-fit temperature of $kT_{\rm dbb}=7.56\pm0.09$~eV.

%The disc component has an average intrinsic luminosity of \red{$XX\ \rm erg\ s^{-1}$} in the $1000$--$7000$~\AA\ band, while the soft-excess luminosity in the $0.2$--$2$~keV range varies between \red{$XX$--$YY\ \rm erg\ s^{-1}$}. The $0.2$--$10$~keV luminosity of the power-law component varies between \red{$XX$--$YY\ \rm erg\ s^{-1}$}. In addition to the primary X-ray continuum, we also included a reprocessed component produced by X-ray reflection, described in the following section.

%-----------------------------------------------------------------
%												SED modelling
%-----------------------------------------------------------------
\begin{deluxetable*}{lcccccc}
\tablecaption{Best-fit parameters of the SED model \texttt{(dbb+comt+pow+refl)*3pion*ebv*reds*2hot}.}
\label{tab:sed}
\tablewidth{0pt}
\tablehead{
\colhead{Parameter} & \colhead{Average} & \colhead{Epoch 1} & \colhead{Epoch 2} & \colhead{Epoch 3} & \colhead{Epoch 4} & \colhead{Epoch 5}
}
\startdata
\multicolumn{7}{c}{\textbf{Power-law (\texttt{pow})}} \\
\hline
Norm.$^{\rm a}$ & $6.18\pm0.03$ & $5.86\pm0.06$ & $2.63\pm0.07$ & $3.31\pm0.08$ & $10.10\pm0.07$ & $6.40\pm0.09$   \\
$\Gamma$   		& $1.95\pm0.01$ & $1.96\pm0.01$ & $1.94\pm0.02$ & $1.98\pm0.02$ & $2.08\pm0.01$  & $1.95\pm0.01$ \\
\hline
\multicolumn{7}{c}{\textbf{Warm Comptonization (\texttt{comt})}} \\
\hline
Norm.$^{\rm a}$        	& $12.7\pm0.2$ 	  & $10.5\pm0.5$ 	  & $4.6\pm0.3$ 	& $4.5\pm0.3$ 	  & $23.6\pm0.7$ 	& $11.0\pm0.05$   \\
$kT_{\rm e}^{\rm c}$ 	& $0.118\pm0.001$ & $0.115\pm0.002$   & $0.117\pm0.003$ & $0.117\pm0.003$ & $0.117\pm0.001$ & $0.123\pm0.001$ \\
$\tau$ 			        & $50^{f}$ 		  & $50^{f}$ 		  & $50^{f}$ 		& $50^{f}$ 		  & $50^{f}$ 		& $50^{f}$ 		  \\
\hline
\multicolumn{7}{c}{\textbf{Disc blackbody (\texttt{dbb})}} \\
\hline
Norm.$^{\rm b}$        	& $1.36\pm0.01$ & $1.31\pm0.2$ & $1.33\pm0.01$ & $1.36\pm0.01$ & $1.38\pm0.02$ & $1.43\pm0.02$ \\
$kT^{\rm c}$     		& $7.56\pm0.09$ & $7.56^{f}$ & $7.56^{f}$ & $7.56^{f}$ & $7.56^{f}$ & $7.56^{f}$ \\
\hline
\multicolumn{7}{c}{\textbf{Cold reflection (\texttt{refl})}} \\
\hline
Scale, $s$       	& $0.88\pm0.3$ 	& $0.89\pm0.07$ & $1.3\pm0.1$ & $1.3\pm0.1$ & $0.75\pm0.06$ & $0.86\pm0.08$ \\
\hline
\multicolumn{7}{c}{\textbf{Galactic absorption (\texttt{hot}, \texttt{ebv})}} \\
\hline
$N_{\rm H, cold}^{\rm d}$   & $3.3^{f}$ & $3.3^{f}$ & $3.3^{f}$ & $3.3^{f}$ & $3.3^{f}$ & $3.3^{f}$ \\
$kT_{\rm cold}^{\rm c}$     & $0.001^{f}$ & $0.001^{f}$ & $0.001^{f}$ & $0.001^{f}$ & $0.001^{f}$ & $0.001^{f}$ \\
$N_{\rm H, warm}^{\rm d}$   & $0.30_{-0.13}^{+0.17}$ & $0.30^{f}$ & $0.30^{f}$ & $0.30^{f}$ & $0.30^{f}$ & $0.30^{f}$ \\
$kT_{\rm warm}^{\rm c}$     & $61_{-10}^{+12}$ & $61^{f}$ & $61^{f}$ & $61^{f}$ & $61^{f}$ & $61^{f}$ \\
$E(B-V)$       	    		& $0.03^{f}$ & $0.03^{f}$ & $0.03^{f}$ & $0.03^{f}$ & $0.03^{f}$ & $0.03^{f}$ \\
\hline
\multicolumn{7}{c}{\textbf{Fluxes$^{\rm e}$}} \\
\hline
$F_{0.3-10\rm\: keV}$   	& $13.2\pm0.1$ 	    & $12.3\pm0.1 $ & $7.15\pm0.2 $ & $8.23\pm0.2 $ & $21.8\pm0.2 $ & $13.8\pm0.2 $ \\
$L_{0.3-10\rm\: keV}$   	& $4.89\pm0.02$ 	& $4.68\pm0.05 $ & $2.16\pm0.06 $ & $2.52\pm0.06 $ & $8.34\pm0.06 $ & $5.53\pm0.08 $ \\
$L_{\rm ion}$   			& $13.5\pm0.1$ 		& $12.3\pm0.5 $ & $7.4\pm0.3 $ & $8.3\pm0.4 $ & $21.8\pm0.5 $ & $13.8\pm0.2 $ \\
\hline
C-stat/dof  				& $2353/1180$ & $1438/1173$ & $1527/1170$ & $1605/1171$ & $1508/1173$ & $1669/1174$ \\
\enddata
\tablecomments{Fixed parameters in the spectral energy distribution analysis are indicated with $(^f)$.\\
%$\tau=50$, $kT_{\rm seed}$ of comptonisation coupled to $kT_{\rm dbb}$, $N_{\rm H, cold}=3.30\times10^{20}$ cm$^{-2}$, $E(B-V)=0.03$. \\
$^{\rm a}$ Units: $10^{51}$ ph s$^{-1}$ keV$^{-1}$ (pow), $10^{53}$ ph s$^{-1}$ keV$^{-1}$ (comt). \\
$^{\rm b}$ Units: $10^{28}$ cm$^{2}$ (dbb). \\
$^{\rm c}$ Units: eV ($kT_{\rm seed}$, $kT_{\rm dbb}$), keV ($kT_{\rm e}$). \\
$^{\rm d}$ Units: $10^{20}$ cm$^{-2}$ \\
$^{\rm e}$ Observed flux in units of $10^{-12}$ erg cm$^{-2}$ s$^{-1}$. Intrinsic X-ray luminosity and ionising luminosity in units of $10^{43}$ erg s$^{-1}$}
\end{deluxetable*}

\subsection{Reflected X-ray continuum}
\label{sec:refl}
The EPIC-pn and \nustar\ spectra of \mrk\ show clear evidence of X-ray reflection, most prominently the Fe~K$\alpha$ emission line at $\sim6.4$~keV and the high-energy Compton hump above 10~keV. To account for this, we included a cold reflection component (\texttt{refl} in \spex; \citealt{Magdziarz95,Zycki94}) that reprocesses the primary X-ray continuum. In our implementation, the reflection spectrum is tied to the same photon index and high-energy cut-off as the observed power-law continuum, while the ionisation parameter was fixed to zero to approximate reflection from neutral material with solar abundances. 

The main free parameter is the reflection scaling factor, which varies modestly ($0.75$--$1.5$) between epochs. This level of variability is plausible, although assuming that the incident continuum is identical to the observed one is a simplification. We note that \texttt{refl} provides a simplified description and does not include relativistic effects from the innermost accretion disc, nor a possible broadened Fe~K$\alpha$ component. A more sophisticated reflection model would be required to capture those features. The model also does not include any contribution from lighter ions (e.g. oxygen, nitrogen) and Fe-L which might also contribute to the soft excess. For the purposes of the present SED modelling, however, this simplified approach is sufficient to reproduce the overall shape of the broadband emission \citep[e.g.][]{Rogantini22a}.

\subsection{Galactic absorption and reddening}
\label{sec:ism} 
In modelling the X-ray spectra of \mrk, we accounted for absorption by the cold, neutral ISM in the Galaxy. This was implemented with the \texttt{hot} component in \spex, which computes the transmitted spectrum of a plasma in collisional ionisation equilibrium. To approximate neutral absorption, we fixed the plasma temperature to $kT = 1\times10^{-6}\ \mathrm{keV}$ ($T\simeq10\ \mathrm{K}$), and adopted a hydrogen column density of $\NH = 3.30\times10^{20}\ \mathrm{cm^{-2}}$ \citep{HI4PI16}.

For the optical/UV data from \swift/UVOT and \xmm/OM (UVW1 filter only), we corrected for Galactic dust reddening using the multiplicative \texttt{ebv} component in \spex. This model applies the extinction curve of \citet{Cardelli89}, with the near-UV update by \citet{ODonnell94}. Based on the recalibration by \citet{Schlafly11}, we adopted $E(B-V)=0.030$~mag and fixed the total-to-selective extinction ratio to the Galactic average, $R_{\rm V}=3.1$.

We also tested for the presence of a warmer Galactic ISM phase by adding a second \texttt{hot} component with temperature left free. The deep RGS spectra (Section~\ref{sec:rgs}) indeed constrain a warmer phase with $kT=61$~eV, which we keep fixed, together with its column density, in the single-epoch SED fits. Finally, we investigated whether intrinsic cold absorption and reddening in \mrk\ could contribute by adding an additional \texttt{hot} and \texttt{ebv} component at the systemic redshift of the source. These components did not improve the fit, indicating that any intrinsic neutral absorption is negligible.

In contrast, the spectra clearly require multiple layers of ionised absorption. Three photoionisation components are necessary to model the ionised outflows in \mrk. These outflows are the main focus of this work and their modeling is described in Section~\ref{sec:photo} and summarised in Table~\ref{tab:outflows}.

\subsection{Emission from the BLR/NLR and galactic bulge}
\label{sec:blr}

The optical/UV photometric bands of \swift/UVOT and \xmm/OM include not only the AGN continuum, but also emission from the broad-line region (BLR) and narrow-line region (NLR). To account for this contribution in the SED modeling, we adopted the empirical emission model of \citet{Mehdipour15a}, originally developed for NGC\,5548, as a template. This model accounts for the Balmer continuum, the Fe\,\textsc{ii} pseudo-continuum, and strong emission lines from the BLR and NLR. For simplicity, we assumed the BLR/NLR emission to be constant across the five \xmm\ epochs. The predicted H$\beta$ flux from our approximate modelling is $\sim5\times10^{-13}\ \ergflux$, consistent with the fluxes reported by \citet{Grier12} during the 2010 campaign, where $F({\rm H}\beta)$ varied between $(5$--$6)\times10^{-13}\ \ergflux$.

The circular extraction apertures of \swift/UVOT (radius $5''$) inevitably include starlight from the host galaxy of \mrk. To correct for this contribution, we adopted the optical bulge flux derived by \citet{Bentz13} from HST imaging. They modelled the stellar emission from the galaxy bulge using a rectangular aperture of $5''\times7.6''$, comparable in size to the UVOT aperture. From their modelling of images taken with the F550M medium-band $V$ filter (exposure of $2040\ \mathrm{s}$), we recalculated the expected flux for our UVOT aperture, obtaining a bulge flux of $1.16\times10^{-15}\ \ergcm$ at 5231~\AA\ with an estimated uncertainty of $\sim10\%$.

To extend this estimate across the optical/UV range, we adopted the galaxy bulge spectral template of \citet{Kinney96}, widely used in previous AGN SED studies \citep[e.g.][]{Mehdipour15a,Mehdipour18}. We scaled the template to match the measured bulge flux at 5231~\AA\ and used it to estimate the stellar-bulge contribution at the wavelengths covered by the OM and UVOT filters. In the top panel of Fig.~\ref{fig:sed}, the contributions of the host bulge and the BLR/NLR emission are shown as dashed pink and purple curves, respectively.

\tabletypesize{\normalsize}

\section{Modeling of the ionised outflow}
\label{sec:photo}

Modeling high-resolution RGS spectra, we find that, in line with past observations, the complex ionised outflow system remains present in the intermediate flux state. The photoionisation modelling of the winds for the stacked spectrum and for the individual epochs is described in the following subsections.

\subsection{The deep RGS spectrum}
\label{sec:rgs}

To characterise the numerous absorption features produced by the ionised outflows in \mrk, we employed the \pion model in \spex\ \citep{Miller15,Mehdipour16}. This self-consistent photoionisation model simultaneously computes the ionisation and thermal balance of the plasma, together with its resulting spectrum, under the assumption of photoionisation equilibrium (PIE). The calculation is based on the SED derived in Section~\ref{sec:sed} from the continuum components of our model using the stacked \xmm, \nustar and \swift data.

%-----------------------------------------------------------------
%													RGS spectrum
%-----------------------------------------------------------------
\begin{figure*}[ht!]
\centering
\includegraphics[width=\hsize]{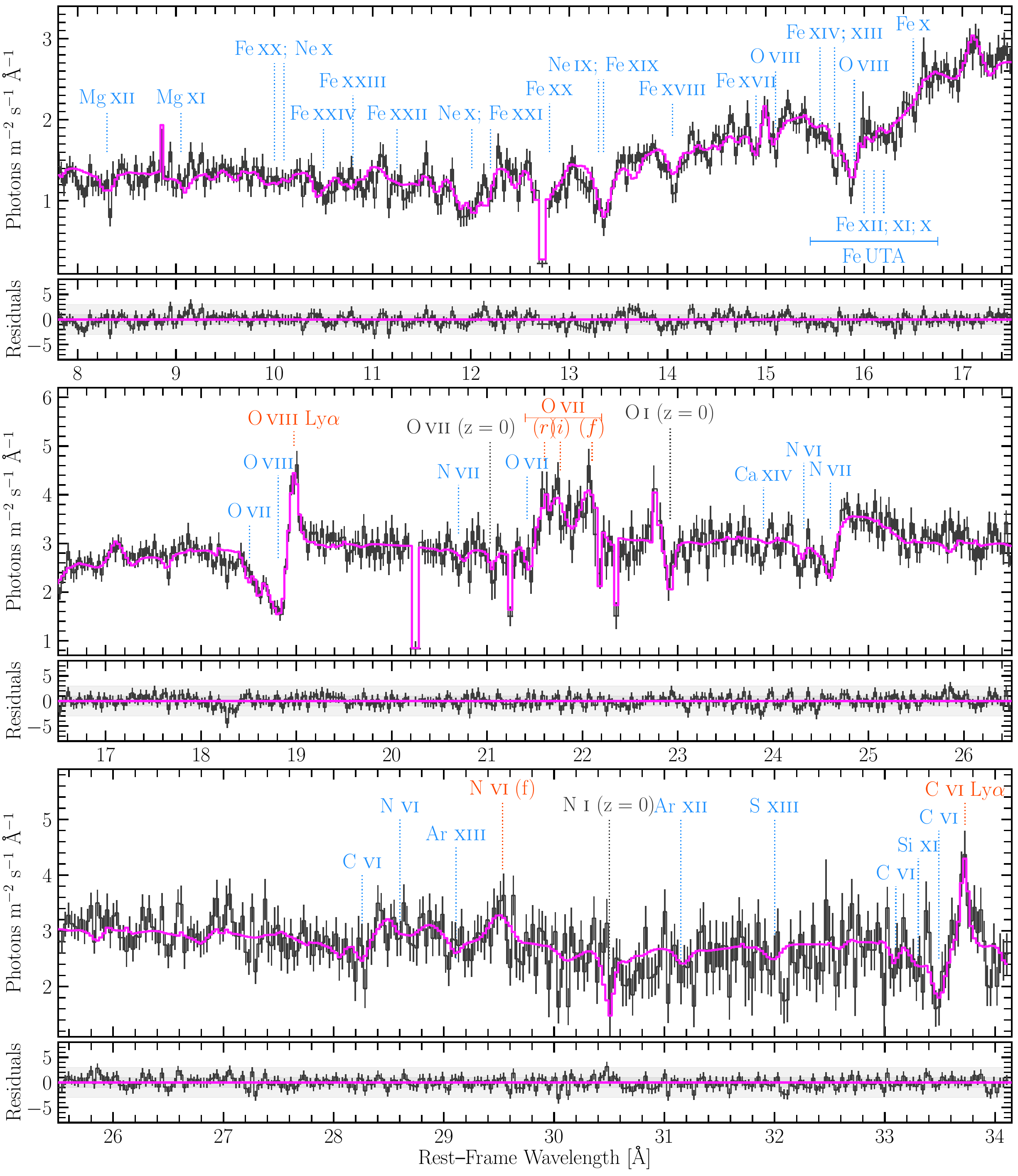}
\caption{Segments of the stacked RGS spectrum of \mrk\ (net exposure of 440~ks), shown in the rest frame of the source.
The magenta line represents the best-fit photoionisation model. Prominent absorption and emission features are marked with blue and red labels, respectively, while Galactic absorption lines at $z=0$ are indicated in black. The absorption features are systematically blueshifted, tracing ionised outflows. Narrow and broad emission lines (e.g.\ \ovii, \oviii, \cvi) are also detected. For display purposes, the RGS1 and RGS2 spectra have been combined in this figure; they were fitted simultaneously in the analysis.}
	\label{fig:rgs}
\end{figure*}

%-----------------------------------------------------------------
%													RGS spectrum
%-----------------------------------------------------------------
\begin{figure*}[ht!]
\centering
\includegraphics[width=\hsize]{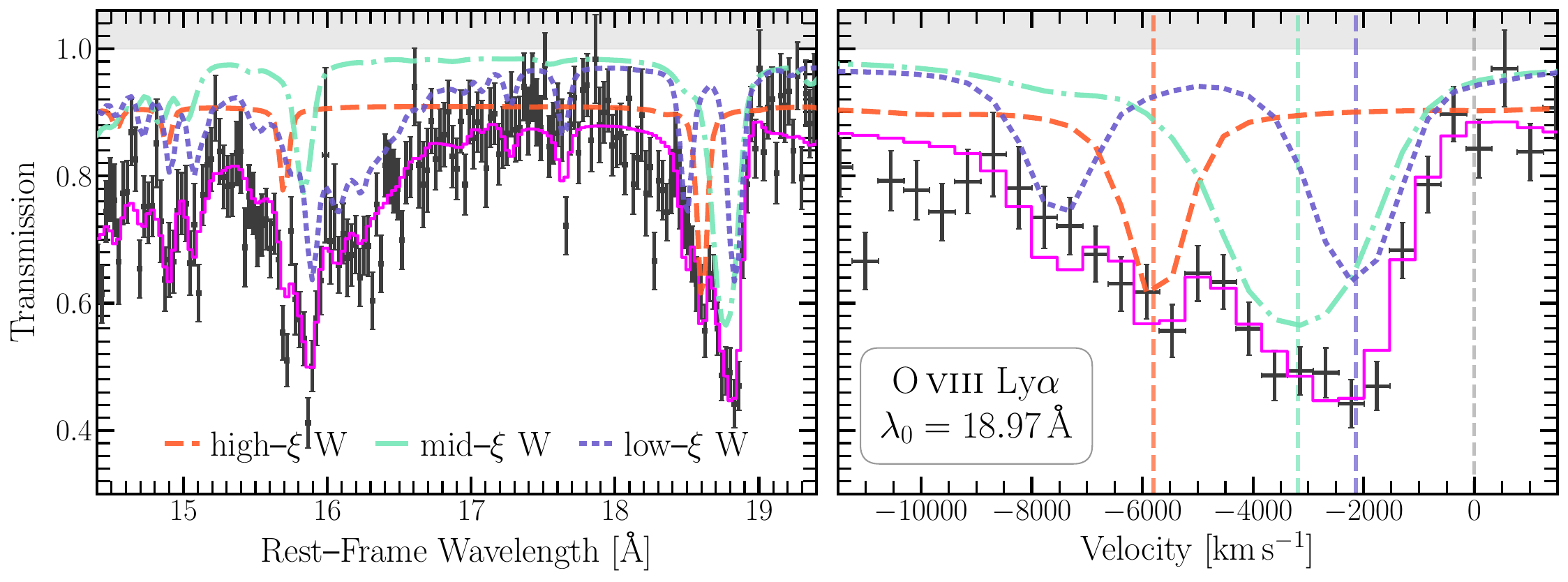}
\caption{\textit{Left panel:} Transmission spectrum of \mrk, obtained by dividing the observed RGS spectrum by the intrinsic continuum model. The plot zooms in on two key regions: the Fe~UTA (15.5--16.5~\AA) and the \oviii\ Ly$\alpha$ line ($\sim$18.7~\AA). The low-$\xi$ wind component provides the dominant contribution to the Fe~UTA, while all three ionisation components are required, in different proportions, to reproduce the \oviii\ absorption profile.
\textit{Right panel:} Zoom-in on the \oviii\ Ly$\alpha$ absorption line, shown in velocity space with respect to the rest-frame wavelength of 18.97~\AA. The black data points and magenta line represent the transmitted RGS spectrum and best-fit model, respectively, while the coloured curves show the transmission of the individual photoionised wind components. The low-, mid-, and high-$\xi$ winds are shown in blue, green, and orange, respectively. Vertical dashed lines mark the best-fit outflow velocities:
$v_{\rm out}=-2100$~km~s$^{-1}$ (low-$\xi$),
$v_{\rm out}=-3200$~km~s$^{-1}$ (mid-$\xi$), and
$v_{\rm out}=-5800$~km~s$^{-1}$ (high-$\xi$).
The figure highlights that the \oviii\ absorption is shaped by the superposition of all three wind components, each contributing with different velocities and strengths.}
	\label{fig:zoom_in}
\end{figure*}

We first obtained the best-fit model for the stacked 2021 RGS spectrum of \mrk\ (net exposure of 440~ks), and then used this model as the baseline for fitting the individual RGS spectra. Figure~\ref{fig:rgs} shows the stacked spectrum in the source rest frame, with the main absorption lines indicated. To reproduce the absorption features from the different ionic species, three distinct \pion components with different ionisation parameters $\xi$ are required \citep{Tarter69}. The ionisation parameter is defined as $\xi = L_{\rm ion}/(n_{\rm H} r^2)$, where $L_{\rm ion}$ is the ionising luminosity in the 1--1000~Ry band (13.6~eV to 13.6~keV) in $\rm erg\ s^{-1}$, $n_{\rm H}$ is the hydrogen density in $\rm cm^{-3}$, and $r$ is the distance between the ionising source and the photoionised plasma in cm. Starting from the continuum model, adding one \pion component improves the fit by $\Delta C=990$ for five additional free parameters. Adding a second and third \pion component yields further improvements of $\Delta C=400$ (five free parameters) and $\Delta C=90$ (four free parameters, with $f_{\rm cov}$ fixed to unity), respectively. The best-fit parameters of the three components are reported in Table~\ref{tab:outflows}. We explored column densities down to $10^{19}~\rm cm^{-2}$, ionisation parameters in the range $\log\xi=-1$ to 5.5, and outflow velocities from a few hundred $\rm km\ s^{-1}$ up to mildly relativistic values; however, adding a fourth component is not statistically required ($\Delta C<10$ for four additional free parameters).

The low-ionisation component (low-$\xi$ wind) accounts for absorption from M-shell Fe ions, producing the broad unresolved transition array (UTA) at $\sim$16--17~\AA\ \citep{Behar01}. It also contributes to transitions from Be-like and Li-like ions such as \sixi, \mgix, and \mgx, as well as the He-like \neix. The intermediate-ionisation component (mid-$\xi$ wind) predominantly produces the He-like \neix\ and \mgxi\ lines, and dominates the H-like transitions of \oviii\ and \nex, together with lines from \fexvii--\fexx. Finally, the high-ionisation component (high-$\xi$ wind) is required to reproduce the strongest H-like \oviii\ and \nex\ absorption, and contributes significantly to highly ionised iron species such as \fexxi, \fexxii, and \fexxiii within the RGS band. Models with fewer components fail to reproduce the full set of detected transitions in the deep, line-rich RGS spectrum.
 
The absorption lines are systematically blueshifted with respect to their rest-frame wavelengths, indicating that the ionised gas is outflowing. We identify three distinct velocity components at $\sim -2100$, $\sim -3200$, and $\sim -5800$~km\,s$^{-1}$, corresponding to the low-, mid-, and high-$\xi$ wind components, respectively. Figure~\ref{fig:zoom_in} (left panel) shows how the low-$\xi$ wind accounts for the UTA feature, and demonstrates that the full three-component ionisation/velocity structure is required to reproduce the observed \oviii\ line profile. The right panel shows the velocity profile of the \oviii\ Ly$\alpha$ absorption line, where the contributions of the three components are evident. Approximately 12\%, 57\%, and 31\% of the total \oviii\ Ly$\alpha$ equivalent width arise from the high-$\xi$, mid-$\xi$, and low-$\xi$ winds, respectively.

\begin{figure*}[t!]
\centering
\includegraphics[width=\hsize]{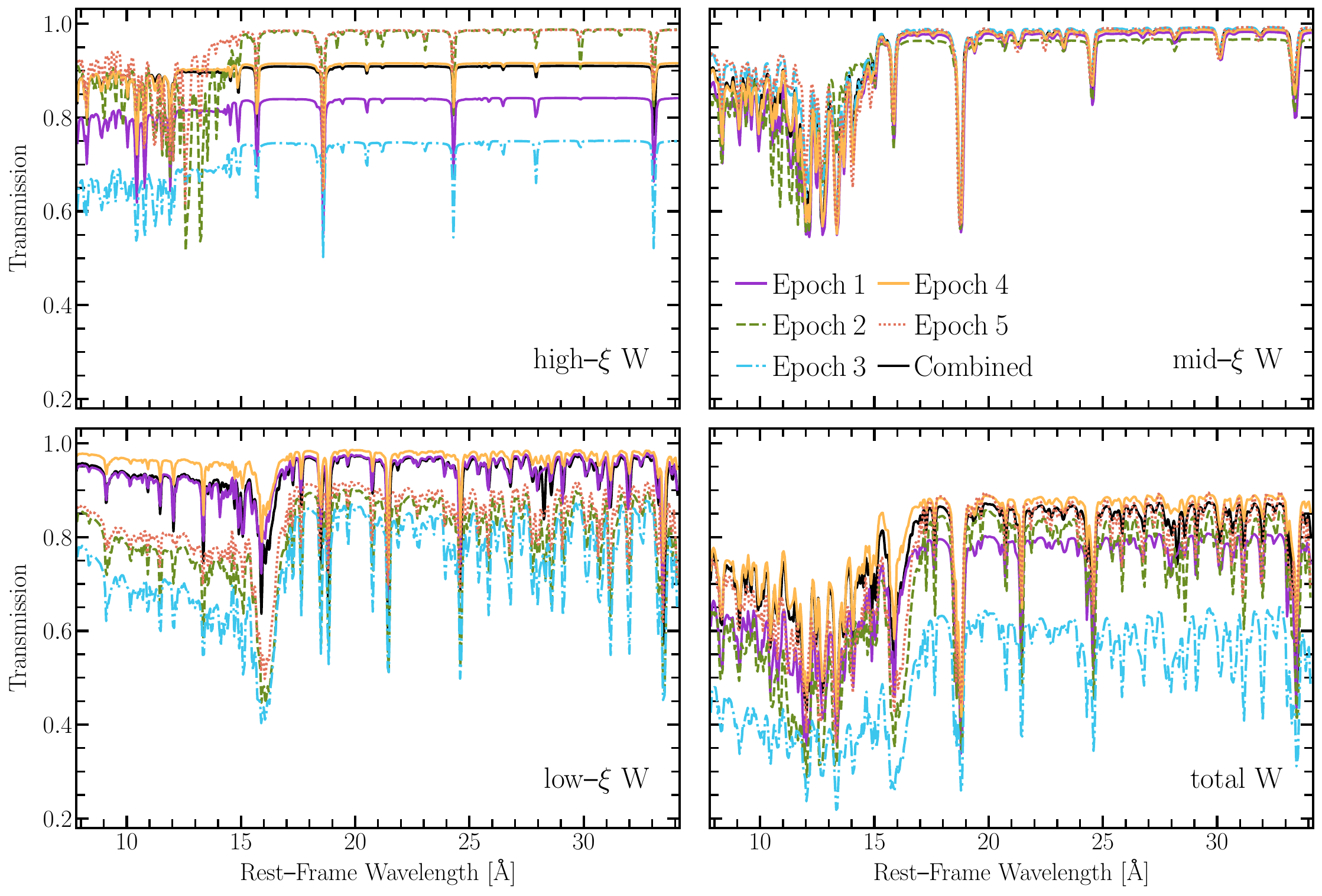}
\caption{
Evolution of the model transmission spectra of the photoionised wind in \mrk.
From top to bottom, the panels show the high-, mid-, and low-ionisation components, and the total transmission (product of all components), for the five \xmm\ epochs (color-coded).
Changes in both the depth and detailed line structure, most prominently in the Fe\,\textsc{xvii}--\oviii\ band, trace variations in the physical state of the outflow.
The low-ionisation component dominates the Fe~UTA region, while the high-ionisation component contributes most strongly to the \oviii\ absorption.}
\label{fig:transmission_evolution}
\end{figure*}

In the time-averaged spectrum, we allowed the column density $N_{\rm H}$ and covering fraction $f_{\rm cov}$ to vary for all three wind components. All three components require substantial columns, with $N_{\rm H}\sim10^{21}$--$10^{23}\ \rm cm^{-2}$ from the low- to the high-ionisation phases. For the high-$\xi$ component, $f_{\rm cov}$ is strongly degenerate with $N_{\rm H}$ because its transmission is close to featureless and is constrained mainly by narrow, high-ionisation absorption lines. This degeneracy can be reduced when the transmission shows appreciable curvature, since curvature driven by the total absorbing column differs from that produced by partial covering. We therefore fixed the high-$\xi$ covering fraction to unity in the stacked-spectrum fit. Figure~\ref{fig:transmission_evolution} shows the model transmission spectra of the three \pion components (solid black lines) for the stacked spectrum, illustrating their respective contributions to the total absorption. The low-$\xi$ component produces most of the broad absorption in the Fe~UTA region, while the high-$\xi$ component contributes primarily to the strongest high-ionisation lines (in particular \oviii). The mid-$\xi$ phase contributes to both regimes and shapes the intermediate-ionisation line forest.

%-----------------------------------------------------------------
%                                    Outflow modelling
%-----------------------------------------------------------------
\begin{deluxetable*}{lcccccc}
\tablecaption{Best-fit parameters of the ionised outflow components obtained with the \pion model.}
\tablewidth{0pt}
\tablehead{
\colhead{Parameter} & \colhead{Average} & \colhead{Epoch 1} & \colhead{Epoch 2} & \colhead{Epoch 3} & \colhead{Epoch 4} & \colhead{Epoch 5}
}
\startdata
\multicolumn{7}{c}{\textbf{High-$\xi$ wind}} \\
\hline
$N_{\rm H}^{\rm a}$    & $12\pm0.9$ & $22\pm0.2$ & $1.4_{-0.8}^{+1.5}$ & $36\pm5$ & $ 11_{-6}^{+14}$ & $1.6_{-0.5}^{+0.9}$ \\
$\log\xi^{\rm b}$      & $3.69_{-0.05}^{+0.06}$ & $3.58_{-0.8}^{+1.0}$  & $3.1\pm0.2$ & $3.51_{-0.05}^{+0.29}$ & $3.72_{-0.11}^{+0.15}$ & $3.39\pm0.07$ \\
$f_{\rm cov}$          & $1^{f}$  & $1^{f}$ & $1^{f}$ & $1^{f}$ & $1^{f}$ & $1^{f}$ \\
$v_{\rm turb}^{\rm c}$ & $130\pm30$  & $120\pm40$ & $130^{f}$ & $130^{f}$ & $ 100\pm30$ & $150_{-50}^{+70}$ \\
$v_{\rm out}^{\rm c}$  & $-5780\pm90$  & $-5760_{-170}^{+150}$ & $-5740_{-160}^{+150}$ & $-5860_{-130}^{+100}$ & $-5640_{-170}^{+100}$ & $-6450_{-160}^{+150}$ \\
\hline
\multicolumn{7}{c}{\textbf{Mid-$\xi$ wind}} \\
\hline
$N_{\rm H}^{\rm a}$    & $2.8_{-0.4}^{+0.5}$  & $3.7\pm0.8$ & $7.6_{-4.4}^{+4.9}$ & $1.5_{-0.5}^{+2.0}$    & $2.6_{-0.4}^{+0.5}$ & $1.1_{-0.3}^{+0.4}$ \\
$\log\xi^{\rm b}$      & $2.97\pm0.03$  & $2.94\pm0.05$ & $3.39_{-0.10}^{+0.06}$ & $2.89\pm0.10$ & $2.94\pm0.04$ & $2.82\pm0.10$ \\
$f_{\rm cov}$          & $0.56\pm0.04$  & $0.56^{f}$ & $0.56^{f}$ & $0.56^{f}$ & $0.56^{f}$ & $0.56^{f}$ \\
$v_{\rm turb}^{\rm c}$ & $730_{-80}^{+90}$ & $810_{-120}^{+180}$ & $730^{f}$ & $730^{f}$ & $790_{-80}^{+110}$ & $900_{-150}^{+160}$  \\
$v_{\rm out}^{\rm c}$  & $-3180\pm100$ & $-2700_{-220}^{+170}$ & $-3200_{-310}^{+270}$ & $-3280\pm260$ & $-2920_{-130}^{+110}$ & $-3590_{-250}^{+210}$ \\
\hline
\multicolumn{7}{c}{\textbf{Low-$\xi$ wind}} \\
\hline
$N_{\rm H}^{\rm a}$    & $0.20\pm0.05$  & $0.23_{-0.06}^{+0.09}$ & $0.67\pm0.09$ & $1.2\pm0.2$ & $0.09\pm0.02$ & $0.64_{-0.09}^{+0.11}$ \\
$\log\xi^{\rm b}$      & $1.91\pm0.02$  & $2.02\pm0.05$ & $1.99\pm0.04$ & $2.00\pm0.03$ & $1.83_{-0.09}^{+0.07}$ & $1.98\pm0.03$ \\
$f_{\rm cov}$          & $0.67_{-0.09}^{+0.14}$ & $0.67^{f}$ & $0.67^{f}$ & $0.67^{f}$ & $0.67^{f}$ & $0.67^{f}$ \\
$v_{\rm turb}^{\rm c}$ & $280\pm60$  & $190_{-60}^{+100}$ & $280^{f}$ & $280^{f}$ & $170_{-40}^{+50}$ & $110\pm40$ \\
$v_{\rm out}^{\rm c}$  & $-2120_{-70}^{+60}$  & $-2480_{-130}^{+160}$ & $-2050_{-100}^{+110}$ & $-2120_{-130}^{+110}$ & $-2100\pm100$ & $-2490_{-120}^{+100}$ \\
\hline
C-stat/dof  & 1780/1462 & 1650/1435 & 1750/1442 & 1676/1443 & 1580/1434 & 1763/1437 \\
\enddata
\tablecomments{Average values represent the best-fit parameters from the stacked RGS spectrum. Fixed parameters are indicated with $(^f)$. For single-epoch fits, $f_{\rm cov}$ was fixed to the stacked-spectrum value (and for the high-$\xi$ component fixed to unity).\\
$^{\rm a}$ Units: $10^{22}\ \rm cm^{-2}$. \\
$^{\rm b}$ Ionisation parameter, $\xi$, in $\rm erg\:cm\:s^{-1}$. \\
$^{\rm c}$ Velocities in km\:s$^{-1}$.}
\label{tab:outflows}
\end{deluxetable*}

%-----------------------------------------------------------------
%													TRANSMISSION
%-----------------------------------------------------------------

We also modelled the turbulent velocities $v_{\rm turb}$, which determine the line broadening. The derived values are $\sim130$--$280$~km~s$^{-1}$ for the low- and high-$\xi$ winds, and significantly larger for the mid-$\xi$ wind, reaching $\sim730$~km~s$^{-1}$.

In addition to the absorption features, the RGS spectrum of \mrk\ shows narrow and broad emission lines. To avoid overloading the fit computation, we modelled these lines phenomenologically with Gaussian profiles. The line identification, centroid energy, normalisation, and width are reported in Table~\ref{tab:emission_lines}. Both the \oviii\ and \cvi\ Ly$\alpha$\ lines require a narrow and a broad component; the narrow and broad components have FWHM of approximately 0.2~\AA\ and 4~\AA, respectively. The \ovii\ triplet is detected through its resonance, intercombination, and forbidden lines. Physical modeling of the emission lines \citep[e.g.][]{Parker19} will be performed in follow-up studies; here they are included simply to ensure high fidelity constraints on the parameters of the ionized absorption models.

\subsection{The variability of the wind components}
\label{sec:tresolved}
To investigate the variability of the three photoionised wind components, we fitted each RGS epoch independently, using the best-fit model of the stacked RGS spectrum as the initial guess. The covering fractions of the three absorbers were fixed to the average values obtained from the stacked spectrum. Given the limited signal-to-noise ratio of the single-epoch spectra, the degeneracy between column density and covering fraction cannot be broken. We also fixed the turbulent velocities in Epochs~2 and~3 to the average values from the stacked fit, since they could not be constrained reliably in these two observations, which have the lowest signal-to-noise due to their low observed flux. For each \pion component, we fitted the column density $N_{\rm H}$, ionisation parameter $\xi$, and outflow velocity $v_{\rm out}$.

For the broadband continuum, we adopted the SED model determined for each observation in the previous step of the analysis (Section~\ref{sec:sed}). The photon index $\Gamma$ and normalisation of the power-law component, as well as the temperature and normalisation of the Comptonisation component, were allowed to vary within $2\sigma$ of their best-fit values from the SED modelling. This allows a consistent adjustment of the continuum within the RGS band while accounting for the detailed absorption. During the fits to the individual RGS spectra, the normalisations of the emission lines were allowed to vary within $2\sigma$ of the best-fit values from the stacked spectrum, while all other Gaussian parameters were kept fixed. This implicitly assumes that these parameters are constant across epochs and that any apparent changes are consistent with their statistical uncertainties.

\begin{figure*}[ht!]
\centering
\includegraphics[width=.8\hsize]{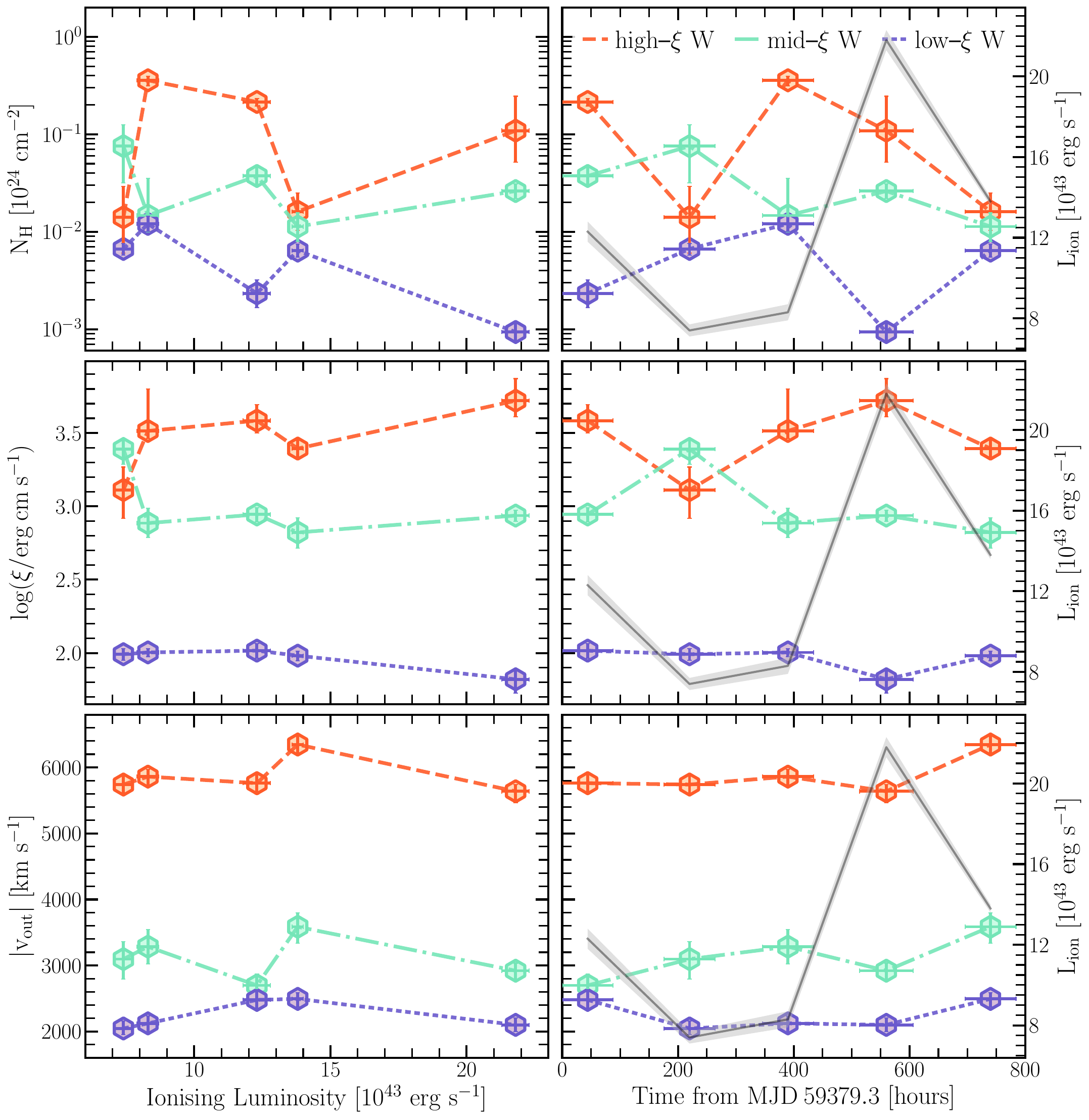}
\caption{Evolution of the ionised outflow parameters of \mrk\ across the 2021 campaign. 
\textit{Left column:} wind parameters as a function of ionising luminosity $L_{\rm ion}$ (13.6~eV--13.6~keV).
From top to bottom, the panels show the column density ($N_{\rm H}$), the ionisation parameter ($\log\xi$), and the absolute outflow velocity ($|v_{\rm out}|$).
\textit{Right column:} same parameters as a function of time since the start of the first observation. The shaded gray band indicates the temporal evolution of the ionising luminosity.
Each point corresponds to one of the five \xmm\ epochs, and the curves connect measurements of the same wind component.
The orange, green, and blue series represent the high-, mid-, and low-$\xi$ wind components, respectively.}
\label{fig:outflow_evolution}
\end{figure*}

Figure~\ref{fig:transmission_evolution} illustrates how the transmission spectra of each component vary across the observations. The differences between epochs show that the absorber opacity does not change as a simple normalisation: both the depth and the detailed line structure vary, consistent with changes in $N_{\rm H}$ and $\xi$ (and, to a lesser extent, in the velocity field). The best-fit parameters of the ionised outflow components for each RGS observation are listed in Table~\ref{tab:outflows}. Figure~\ref{fig:outflow_evolution} shows the evolution of the column density, ionisation parameter, and outflow velocity as a function of time (left panels) and ionising luminosity (right panels). Zoomed-in RGS spectra for each epoch are shown in Appendix~\ref{sec:appendix} (Fig.~\ref{fig:rgs_evolution}).

%-----------------------------------------------------------------
%													RGS spectrum
%-----------------------------------------------------------------

%-----------------------------------------------------------------
%                                    Emission lines
%-----------------------------------------------------------------
\begin{table*}
\caption{X-ray emission lines detected in the stacked RGS spectrum of \mrk.}
\label{tab:emission_lines}
\centering
\begin{tabular}{lcccc}
\hline\hline
\noalign{\vskip 0.75mm}
Line ID & Wavelength & Reference wavelength & Norm & FWHM \\
        & (\AA) & (\AA) & ($10^{49}\,\rm ph\,s^{-1}$) & (\AA) \\
\noalign{\vskip 0.75mm}
\hline
\noalign{\vskip 0.75mm}
C\,\textsc{vi} (broad)      & $34.641_{-0.817}^{+4.940}$ & 33.738  & $88.12_{-25.74}^{+41.88}$ & $4.675_{-1.623}^{+6.559}$ \\
C\,\textsc{vi} Ly$\alpha$   & $33.724 \pm 0.017$         & 33.738  & $4.91 \pm 0.93$           & $<0.161$ \\
N\,\textsc{vi} ($f$)        & $29.496 \pm 0.035$         & 29.552  & $3.74 \pm 0.96$           & $0.215_{-0.046}^{+0.060}$ \\
N\,\textsc{vi} ($i$)        & $28.870 \pm 0.060$         & 29.080  & $2.08 \pm 0.83$           & $0.215^{c}$ \\
N\,\textsc{vi} ($r$)        & $28.517 \pm 0.050$         & 28.781  & $3.12 \pm 0.85$           & $0.215^{c}$ \\
N\,\textsc{vii} Ly$\alpha$  & $24.890 \pm 0.038$         & 24.791  & $6.83 \pm 0.88$           & $0.505 \pm 0.086$ \\
O\,\textsc{vii} ($f$)       & $22.062 \pm 0.021$         & 22.105  & $6.29 \pm 0.91$           & $0.182_{-0.021}^{+0.025}$ \\
O\,\textsc{vii} ($i$)       & $21.741 \pm 0.021$         & 21.799  & $5.62 \pm 0.78$           & $0.182^{c}$ \\
O\,\textsc{vii} ($r$)       & $21.487 \pm 0.022$         & 21.603  & $4.58 \pm 0.73$           & $0.182^{c}$ \\
O\,\textsc{viii} (broad)    & $18.874 \pm 0.160$         & 18.965  & $23.94 \pm 2.34$          & $3.310 \pm 0.331$ \\
O\,\textsc{viii} Ly$\alpha$ & $18.975 \pm 0.005$         & 18.965  & $3.58 \pm 0.33$           & $0.063 \pm 0.015$ \\
Fe\,\textsc{xvii}           & $17.081 \pm 0.018$         & 17.055  & $2.31 \pm 0.47$           & $0.154_{-0.038}^{+0.071}$ \\
O\,\textsc{vii} RRC         & $16.679 \pm 0.035$         & 16.771  & $1.51 \pm 0.45$           & $0.210 \pm 0.100$ \\
\hline
\end{tabular}
\tablecomments{Line widths are in \AA. We tied ($^c$) the FWHM of the forbidden, intercombination, and resonance lines of the N\,\textsc{vi} and O\,\textsc{vii} triplets, since they cannot be separated reliably. Reference wavelengths are taken from \citet{Hell25} and references therein. ``RRC'' denotes a radiative recombination continuum, i.e.\ an emission edge produced when free electrons recombine into a bound level; the quoted reference wavelength for O\,\textsc{vii} RRC corresponds to the recombination edge (ionisation threshold) and is taken from \citet{Kaastra08}.}
\end{table*}

\section{Discussion} \label{sec:discussion}
Our deep \xmm\ campaign captured \mrk\ in an intermediate-flux state, revealing three photoionised absorption components associated with an obscuring outflow. The five consecutive observations enable, for the first time, a detailed characterisation of the short-timescale, day-scale variability of a multi-phase obscurer in this source. Previous studies of obscurers in Seyfert~1 galaxies have typically focused either on a single long observation \citep[e.g.][]{Zaidouni24}, on widely separated epochs \citep[e.g.][]{Mao22b}, or on events so optically thick that the RGS spectrum was strongly suppressed \citep[e.g.][]{Kaastra14,Kara21}. The 2021 campaign therefore provides a rare opportunity to study both the structure of the obscurer and its response to continuum variability on timescales comparable to the expected dynamical and ionisation timescales of BLR-scale gas. At the same time, \mrk\ has shown strong long-term X-ray variability over the past two decades, moving between high-flux, weakly absorbed states, intermediate states with prominent ionised absorption, and deep low-flux states dominated by emission lines. To place the 2021 campaign in this broader context, we first summarise the key historical observations and previous results on the source. We then discuss the short-timescale variability of the 2021 obscurer, its physical structure, location, and energetics.

\subsection{The obscuring wind in Mrk~335: from long-term evolution to our 2021 campaign} \label{sec:longterm}
Across 15 \xmm\ observations spanning $\sim$20~years (Dec~2000--Jul~2021; total exposure $\sim$1.4~Ms), \mrk\ displays dramatic transitions between spectral states, from bright, weakly absorbed epochs to faint states in which the continuum is strongly suppressed and the spectrum becomes line dominated. Proposed explanations include intrinsic changes in the coronal continuum \citep{Parker14b,Wilkins15a,Gallo18b} and variable absorption along the line of sight \citep{Komossa20}. Disentangling these scenarios requires more accurate continuum modelling, including complex state-of-the-art reflection models and multi-component low-ionisation/neutral partial-covering absorption. In the following, we focus on the wind and obscuration properties in the broader context of the long-term X-ray/UV behaviour, while remaining agnostic about the physical driver of the observed variability.

\subsubsection{From high-flux states to the onset of obscuration (2000--2007).}
During the first three \xmm\ observations in 2000 and 2006, \mrk\ was observed in a high-flux, weakly absorbed state, with a $0.3$--$10$~keV flux of $\sim5\times10^{-11}\ \ergflux$ \citep{Longinotti07,ONeill07}. In these spectra, only a tenuous ionised outflow was detected, with $\log\xi\sim1.2$ and $\NH\sim10^{19}\ \rm cm^{-2}$, although its velocity changed substantially between epochs, from $\sim2500~\kms$ in 2000 to $\sim7500~\kms$ in 2006 \citep{Liu21}. This behaviour changed dramatically in 2007, when \swift\ caught \mrk\ in a low-flux state lasting at least two months, with the observed X-ray flux dropping by about an order of magnitude \citep{Grupe07}. The contemporaneous RGS spectrum revealed prominent soft X-ray emission lines \citep{Longinotti08}, and the spectral changes were mainly concentrated below 2~keV. Both a reflection-dominated scenario and a partial-covering model with compact clumps obscuring the X-ray source could reproduce the 2007 spectrum \citep{Grupe08}. Photoionisation modelling later identified a low-ionisation absorber with $\log\xi\sim0.7$, $\NH\sim2\times10^{20}\ \rm cm^{-2}$, and $\vout\sim-6500~\kms$ \citep{Liu21}.

\subsubsection{Intermediate-flux state and emergence of a multi-phase obscurer (2009).}
Two consecutive \xmm\ orbits in 2009 (total exposure of $\sim200$~ks) captured \mrk\ in an intermediate-flux state ($F_{0.3-10~\rm keV}\sim10^{-11}\ \ergflux$) and revealed a complex ionised wind \citep{Longinotti13}. The high-resolution grating spectra required three distinct ionised components, all outflowing at relatively high velocities of $(4$--$6)\times10^{3}~\kms$. Notably, the presence of a variable, complex ionised absorber had already been proposed to explain the peculiar 1987 \ginga\ spectrum \citep{Turner93}, for which an ionised absorber with $\NH\sim10^{22}\ \rm cm^{-2}$ was invoked.

\subsubsection{Deep low-flux states and line-dominated spectra (2015--2019).}
The 2009 intermediate-state observation was followed by four \xmm\ Target-of-Opportunity observations (2015, 2018, and early and late 2019) that captured \mrk\ in its lowest X-ray flux states \citep{Longinotti19,Parker19,Liu21}. During these epochs, the grating spectra are dominated by strong photoionised emission lines, together with an ionised absorber characterised by $\log\xi \approx 1.3$--1.5, $\NH \approx 10^{20\text{--}21}\ \rm cm^{-2}$, and $\vout \approx -5700~\kms$. %To model the emission features, \citet{Parker19} and \citet{Liu21} employed photoionisation calculations using \texttt{pion\_xs}, a table implementation of the \pion\ model within \textsc{xspec}. 
For the 2018--2019 observations, two emission components ($\log\xi \sim 1.0$ and 2.4, with $n_{\rm H}\sim10^{6}\ \rm cm^{-3}$) were required to reproduce the rich line forest \citep{Parker19}. The higher-ionisation wind component was not detected during these low-flux epochs; it may nevertheless be present, with the limited signal-to-noise ratio preventing a robust detection. \citet{Liu21} found that the emission components vary in ionisation state and flux across epochs, and are blueshifted with respect to the systemic frame, with velocities ranging from $\sim$100~km~s$^{-1}$ in observations after 2015 up to a few thousand km~s$^{-1}$ in earlier epochs.

The \ovii\ triplet displayed dramatic changes in line ratios between 2018 and 2019: in 2018 the resonance line dominated, whereas in 2019 the forbidden line became the strongest. \citet{Parker19} interpreted this change as the result of a lower ionisation state and stronger attenuation of the resonance line, implying an extended scattering/emitting region ($r \gtrsim 0.1$~pc) rather than BLR-scale gas.
By contrast, the \ovii\ triplet in the 2007 low-flux observation was dominated by the intercombination line, indicating much higher densities \citep[$n_{\rm e}\sim10^{10\text{--}11}\ \rm cm^{-3}$;][]{Longinotti08}. This points to an origin on BLR scales, comparable to the spatial scale inferred for the obscuring wind in the 2009 intermediate-flux observation \citep{Longinotti13} and in the 2021 observations analysed here.

\subsubsection{The obscurer during our 2021 campaign}
\begin{figure*}[t]
\centering
\includegraphics[width=0.9\hsize]{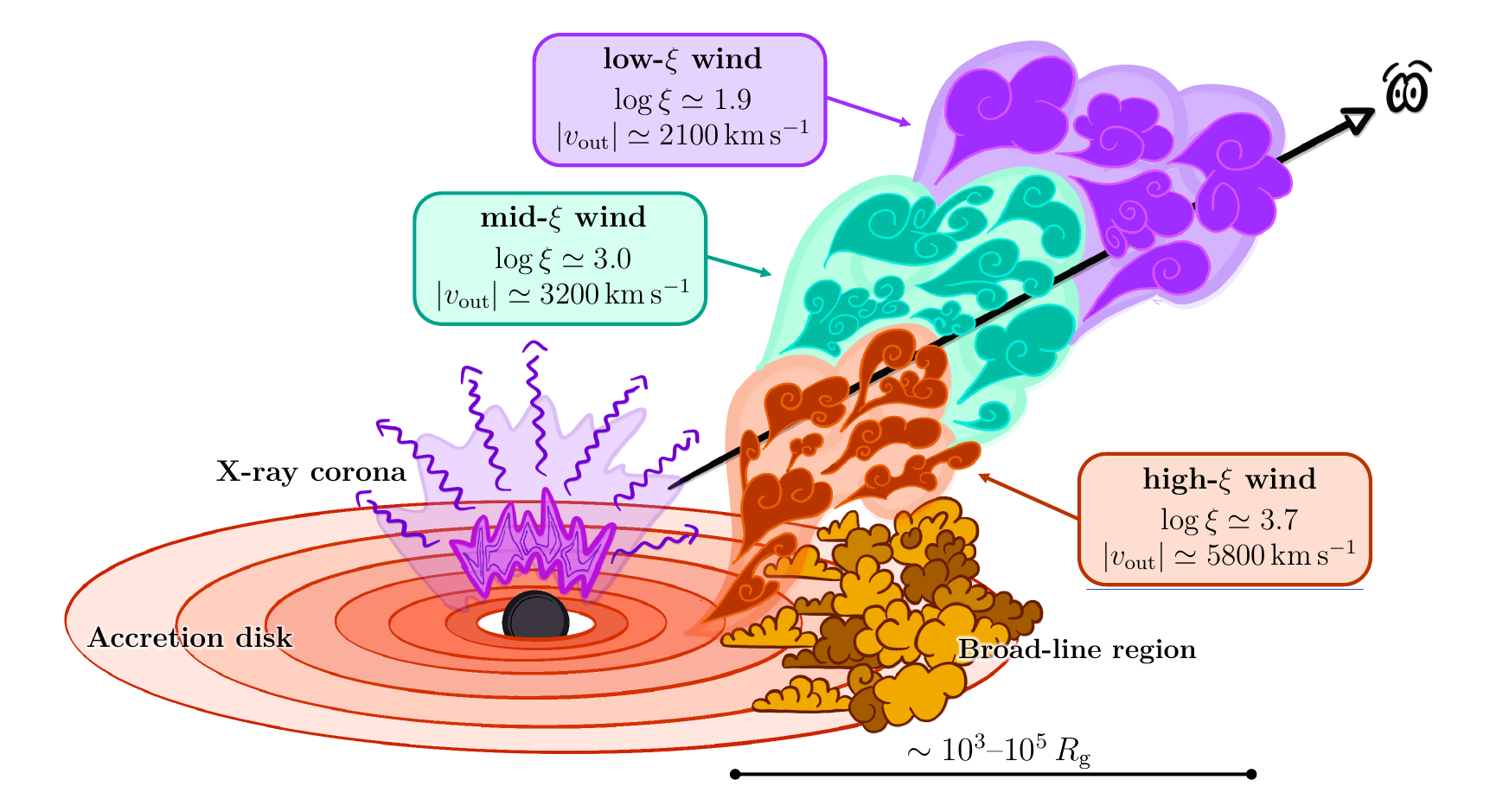}
\caption{Schematic view of the clumpy, multi-phase obscuring wind in Mrk 335.
The obscurer is illustrated as a stratified disk wind intersecting our line of sight. The high-ionisation phase is shown closer to the X-ray corona, while the mid- and low-ionisation phases trace progressively more shielded regions of the same outflow at possible broad-line-region scales, $\sim 10^{3}$--$10^{5},R_{\rm g}$. The sketch is not to scale.}
\label{fig:sketch}
\end{figure*}

During the intermediate-flux phase, the RGS spectra of Mrk\,335 reveal three distinct photoionised absorbers with time-averaged ionisation parameters of $\log \xi \simeq 3.69$ (high-$\xi$), $2.97$ (mid-$\xi$), and $1.91$ (low-$\xi$). The absorption lines are blueshifted by $v_{\rm out} \approx 5800~\mathrm{km~s^{-1}}$, $3200~\mathrm{km~s^{-1}}$, and $2100~\mathrm{km~s^{-1}}$, respectively. These velocities are considerably higher than those typically found for classical warm absorbers ($\sim$100--1000~$\mathrm{km~s^{-1}}$; \citealt[e.g.][]{Blustin05,Detmers11,Laha14,Behar17}), and lower than the velocities attributed to ultra-fast outflows (UFOs), which can reach a substantial fraction of the speed of light \citep[e.g.][]{Tombesi10b,Xrism25a}. Instead, the velocities are comparable to the absorption troughs seen in the UV band during episodic obscuration events \citep[e.g.][]{Kaastra14,Zaidouni24}. A schematic interpretation of the multi-phase obscurer during the 2021 campaign is shown in Fig.~\ref{fig:sketch}.

%-----------------------------------------------------------------
%											              Sketch
%-----------------------------------------------------------------

The physical parameters of the 2021 obscurer are broadly consistent with the ionisation states ($\log\xi \sim 1.8,\,2.1,\,3.3$) and column densities ($\NH \sim 10^{21\text{--}22}\ \rm cm^{-2}$) measured during the 2009 campaign, suggesting that a similar obscuring phenomenology can persist in \mrk\ across decade timescales. The ionisation parameters reported in \citet{Longinotti13} are slightly lower than in 2021, which may reflect differences in the assumed SED and/or photoionisation modelling choices (see e.g.\ \citealt{Mehdipour16} for the impact of SED and code differences). Nevertheless, their transmission spectra show the same qualitative requirements: a low-ionisation component is needed to reproduce the Fe~UTA, and multiple components are required to fit high-ionisation lines such as \oviii. In 2009 the outflow velocities of the low- and mid-ionisation phases were higher ($\sim$4000 and $\sim$5100 km/s respectively) than those we measure in 2021 for the corresponding components, which may indicate intrinsic variability in the wind kinematics and is consistent with the short-term velocity changes we observe during the 2021 flare.

\subsubsection{UV connection and multiwavelength perspective.}

Although no simultaneous \hst/COS spectra are available for our 2021 campaign, previous high-resolution UV observations between 1994 and 2018 have revealed strong, broad, blueshifted absorption in Ly$\alpha$ and \civ\ \citep{Longinotti13,Longinotti19,Parker19}. In 2009, the UV absorption troughs ($\vout \sim -6500~\kms$, $FWHM\sim2600~\kms$) had velocities comparable to the X-ray absorption features in the contemporaneous intermediate-flux \xmm\ data. \citet{Longinotti13} also reported substantial variability of the \civ\ absorption on month-to-year timescales, consistent with column-density changes driven by bulk motion, underscoring the diagnostic power of combining high-resolution UV and X-ray spectroscopy.

Two quasi-simultaneous \hst/COS and \xmm\ campaigns were performed in 2015/2016 and 2018. In 2015/2016, the outflow velocities inferred in the two bands were remarkably consistent ($v_{\rm out,\,X}\sim5700~\kms$ and $v_{\rm out,\,UV}\sim5200~\kms$, with $FWHM\sim700~\kms$; \citealt{Longinotti19}). While the soft X-ray absorber covered nearly the entire X-ray source ($C_{\rm f,X}\approx1$), the UV absorber covered only $\sim20$--30\% of the ionising continuum, suggesting either a more extended UV-emitting region and/or strong inhomogeneity in the absorber. In 2018 the broad UV absorption remained visible, but its velocity decreased to $\sim3000~\kms$, closer to the velocities of the low-$\xi$ and mid-$\xi$ wind components observed in 2021, while the X-ray outflow maintained a higher velocity \citep{Parker19}, indicating that the UV and X-ray absorbing zones can evolve differently even when they are likely related.

\subsubsection{Implications: a long-lived, clumpy obscurer.}
 %Meanwhile, the observed X-ray state is regulated by both (i) changes in the line-of-sight column density and covering fraction of a clumpy outflow and (ii) variations in the intrinsic luminosity, which also modulate the ionisation state of the wind components. 
%During low state XMM observations, only the low-xi wind component is detected, the signal to noise is too low to explore the presence of the other two componetns, which are detected during intermediate flux phases. 
%In this framework, the extended low-flux intervals seen in the \swift\ light curve (Fig.~\ref{fig:pn_nu_lc}) may correspond to episodes in which dense clumps intercept the line of sight and suppress the soft X-ray continuum, whereas high states occur when the effective obscuration is reduced (through higher ionisation, smaller covering fraction, and/or geometric displacement of clumps out of the line of sight) \citep{Grupe07,Komossa20}. At the same time, the intrinsic source varies substantially across the years \citep{Wilkins15a,Gallo15}: during the 2021 campaign, the best-fit SED changes from epoch to epoch (Table~\ref{tab:sed}), indicating that intrinsic continuum variability and variable obscuration act together to shape the observed long-term behaviour.

The long-term behaviour of \mrk\ therefore supports a picture in which a multi-phase, moderately fast obscuring wind is an intrinsic component of the accretion flow, persistent on decade timescales rather than arising from a transient external absorber. The similarity between the 2009 and 2021 intermediate-state spectra suggests that the wind can maintain a broadly similar ionisation/velocity stratification over $\gtrsim 10$~yr. During the lowest-flux \xmm\ observations, only the low-$\xi$ component is detected \citep{Liu21}; the limited signal-to-noise and the strong suppression of the soft continuum prevent a meaningful search for weaker, higher-ionisation features. The absence of the mid- and high-$\xi$ phases in these spectra therefore does not necessarily imply that they are physically absent. Instead, a simple interpretation is that the same multi-phase outflow persists, while variations in the line-of-sight covering fraction and/or total column density---and, potentially, changes in the intrinsic continuum level---modulate both the observed flux and the detectability of the more highly ionised components. This is consistent with the intermediate-state epochs, where the higher-ionisation phases re-emerge as the continuum recovers.

A natural implication is that this multi-phase wind contributes to the long-term obscuration behaviour of \mrk. The long-term \swift/XRT light curve (Fig.~\ref{fig:pn_nu_lc}) shows that the source spends extended periods in suppressed X-ray states, punctuated by occasional recoveries to higher flux. Variations in the column density and/or covering fraction of a clumpy obscurer crossing the line of sight, while the outflow kinematics remain broadly stable, could account for part of the observed X-ray variability. The similarity between the X-ray and UV outflow velocities further supports an origin at BLR scales, where changes in the projected covering fraction on timescales of days to months are expected.

\subsection{Short-term Variability of the Obscuring Wind} \label{sec:short_term}
The five consecutive \xmm\ pointings provide a unique opportunity to probe the variability of the obscuring wind on day timescales. The best-fit wind parameters for each epoch are reported in Table~\ref{tab:outflows}. To search for correlations, Fig.~\ref{fig:outflow_evolution} shows the column density, ionisation parameter, and outflow velocity of all three components as a function of time and of the ionising luminosity $L_{\rm ion}$, integrated over 1--1000~Ry.

The column densities vary significantly around their time-averaged values. The most prominent change is the drop in $N_{\rm H}$ of the low-$\xi$ component during the flare in Epoch~4. We note that the covering factors were kept fixed during the single-epoch fits. Allowing both $N_{\rm H}$ and $f_{\rm cov}$ to vary leads to a significant degeneracy between the two parameters, especially in the single-epoch spectra where the signal-to-noise ratio is lower. Two-dimensional confidence contours show that larger column densities can be partially compensated by smaller covering factors, and vice versa. However, one-dimensional $C$-stat profiles, obtained by stepping each parameter individually, show that $N_{\rm H}$ has a more clearly defined minimum. In contrast, the profile for $f_{\rm cov}$ remains relatively flat toward high covering fractions and does not provide a closed upper confidence bound within the allowed range $0\leq f_{\rm cov}\leq1$. This suggests that $N_{\rm H}$ is better constrained than $f_{\rm cov}$, although the two-dimensional contours confirm that the two parameters remain significantly degenerate. We therefore interpret the Epoch~4 opacity change primarily in terms of a change in $N_{\rm H}$, under the assumption that the covering factor remains constant. On day timescales, such $N_{\rm H}$ variations can be produced by a clumpy absorber moving transversely across the line of sight and/or by changes in the wind structure during high-flux phases. In particular, an enhanced radiation field could reduce the opacity of the lowest-ionisation phase by changing its ionisation state and/or by accelerating absorbing clumps out of the line of sight. A second drop in column density is seen for the high-$\xi$ component during Epoch~2. In this case, however, the additional degeneracy between $N_{\rm H}$ and $\xi$, together with the lower signal-to-noise ratio, limits the robustness of the inferred change.
 
Interestingly, the flare in Epoch~4 is followed by an increase in outflow velocity in all three components during Epoch~5. This trend suggests a possible connection between the enhanced radiation field and the subsequent change in wind kinematics. The acceleration appears to scale with ionisation state, being largest for the high-$\xi$ component and smallest for the low-$\xi$ phase. In the same picture, the increased radiation field during the flare can temporarily reduce the opacity of the coldest phase and lower its line-of-sight column density accelerating for instance the absorbing clumps out of the line of sight.

A qualitatively similar behaviour was reported for Mrk~817, where the obscurer column density reached a minimum during a flaring phase in the long monitoring campaign \citep{Partington23}. In that case, the available CCD-resolution data did not allow a meaningful test for velocity changes. Our RGS measurements therefore provide an important observational link between a flare-driven decrease in obscuration and a subsequent increase in outflow velocity, supporting a scenario in which continuum variability may influence the opacity and kinematics of the obscurer on day timescales.

During Epoch~2, the high-$\xi$ and mid-$\xi$ components exchange their ionisation parameters: the mid-$\xi$ phase reaches values comparable to the average high-$\xi$ state, while the high-$\xi$ component attains its minimum $\log\xi$. This occurs simultaneously with a decrease in $N_{\rm H}$ of the high-$\xi$ component and an increase in $N_{\rm H}$ of the mid-$\xi$ component, while the outflow velocities remain consistent with their average values within uncertainties. We investigated whether this behaviour could be an artefact of local minima by repeating the fit with a grid of initial conditions for $(N_{\rm H},\xi)$ and by allowing the covering fractions to vary. None of these tests yielded a statistically significant improvement, supporting the solution reported in Table~\ref{tab:outflows}. Nevertheless, Epoch~2 also corresponds to the lowest observed X-ray flux of the campaign, and both $N_{\rm H}$ and $\xi$ are consequently more weakly constrained.

In the top-right panel of Fig.~\ref{fig:outflow_evolution}, the high-$\xi$ component shows an approximate positive correlation between $\log\xi$ and $L_{\rm ion}$, indicating that the gas responds to changes in the ionising continuum on timescales of a few days. This is expected if the high-$\xi$ phase is sufficiently close to the central engine. This interpretation is further supported by the temporal evolution shown in the top-left panel, where the ionisation state of the high-$\xi$ component broadly follows the same trend as the ionising luminosity. The observed change between the lowest and highest flux states, $\Delta\log\xi = 0.62_{-0.23}^{+0.25}$, is consistent with the variation expected from the measured change in $L_{\rm ion}$ (a factor of $\sim3$ over the campaign), suggesting that the gas remains close to photoionisation equilibrium. If the response time is shorter than $\sim1$~day (comparable to the observation spacing), the implied recombination timescale requires densities $n \gtrsim 5\times10^{6}\ \mathrm{cm^{-3}}$ and distances $\lesssim0.03$~pc (i.e.\ $\sim2\times10^{4}\ R_{\rm g}$ for $M_{\rm BH}\simeq2.7\times10^{7}\ M_{\odot}$; \citealt{Grier12}). These estimates are order-of-magnitude and depend on the dominant ions controlling the recombination time and on the assumed SED.

For the mid-$\xi$ component, any trend with $L_{\rm ion}$ is less conclusive, largely due to the outlying point during the faintest observation (Epoch 2) and the associated large uncertainties. The low-$\xi$ component does not show a clear instantaneous correlation with $L_{\rm ion}$, but the time evolution suggests a possible delayed response of order $\sim1$~day: $\log\xi$ decreases during Epoch~4 after the lower ionising flux in Epoch~3 and increases again in Epoch~5 following the flare. A one-day lag has also been inferred for the obscurer in Mrk~817 from high-cadence \nicer\ monitoring \citep{Partington23}. If such a delay is real in \mrk, it would place the low-$\xi$ phase further out than the high-$\xi$ component, at distances of order $\sim0.1$~pc ($\sim10^{5}\ R_{\rm g}$). This inference remains tentative, however, because the gaps between observations limit our knowledge of the full ionising-flux history, introducing uncertainty in any lag estimate. The sparse sampling also prevents a robust time-dependent photoionisation analysis \citep[e.g.][]{Rogantini22b}, which would otherwise provide direct constraints on the gas density and distance.

Overall, the day-timescale X-ray variability of \mrk\ is driven by a combination of intrinsic continuum changes, i.e.\ variations in the ionising SED, and variable absorption, i.e.\ changes in the transmission spectrum; see also Fig.~\ref{fig:transmission_evolution}. The strong day-scale changes in opacity, shown in the top panels of Fig.~\ref{fig:outflow_evolution}, are interpreted here primarily as changes in column density under the assumption of constant covering factor. This behaviour supports a clumpy outflow geometry.

To compare intrinsic and absorption-driven variability in an empirical way, we combined our simplified SED approach with the RGS modelling of the obscuring wind and decomposed the band-limited observed flux as $F_{\rm obs}=F_{\rm int}\,T$, where $F_{\rm int}$ is the unabsorbed continuum flux in our baseline description and $T\equiv F_{\rm obs}/F_{\rm int}$ is the corresponding effective transmission. Here, $T$ quantifies only the attenuation produced by the three photoionised wind components included in our RGS model; it does not include any additional neutral or low-ionisation partial-covering absorption that may be required in alternative broadband interpretations. This decomposition is intended purely as an empirical diagnostic to separate changes in the baseline continuum from line-of-sight attenuation within our adopted parametrisation. It does not, by itself, identify the physical origin of the variability, which may involve intrinsic coronal changes, relativistic reflection, and/or partial-covering absorption. Between two epochs $a$ and $b$, this gives $\Delta\ln F_{\rm obs}=\Delta\ln F_{\rm int}+\Delta\ln T$. In the 0.3--10~keV band, $F_{\rm int}$ varies by a factor of $\simeq3.3$ across the epochs, spanning $F_{\rm int}\simeq(0.91$--$3.00)\times10^{-11}\ {\rm erg\,cm^{-2}\,s^{-1}}$. Over the same period, $T$ spans $\simeq0.54$--$0.80$, corresponding to a factor of $\simeq1.5$. Within this parametrisation, changes in the baseline continuum level therefore dominate the broadband variability, while attenuation by the modelled wind provides an additional, non-negligible modulation.

This scenario is also supported by the short-timescale variability seen in the \xmm\ and \nustar\ light curves. In particular, the hardness ratio shown in the bottom panel of Fig.~\ref{fig:pn_nu_lc} reveals several short soft-flare episodes, especially during Epochs~1 and 2. We attempted a finer time-resolved spectroscopic analysis to test whether these events are associated with changes in the obscurer, but the results were limited by the low signal-to-noise ratio of the short-exposure RGS spectra. The \xmm\ and \nustar\ light curves are consistent with a mixed variability scenario, in which changes in the obscuring gas modulate the soft X-ray band, while intrinsic continuum variations contribute to the harder-band variability observed by \nustar\ above 3~keV. A dedicated broadband analysis of the X-ray variability will be presented in future work.

\subsection{Physical structure of the obscurer}
\label{sec:wind_structure}

Having established that the obscurer persists over long timescales and varies on day timescales, we now examine what the fitted wind parameters imply for its physical structure. We focus on the ionisation and velocity stratification of the flow, its thermal stability, its similarity to other Seyfert obscurers, and whether the absorption is better described by photoionised or collisionally ionised gas.

\subsubsection{AMD and ionisation--velocity stratification}

\begin{figure*}[t]
\centering
\includegraphics[width=\hsize]{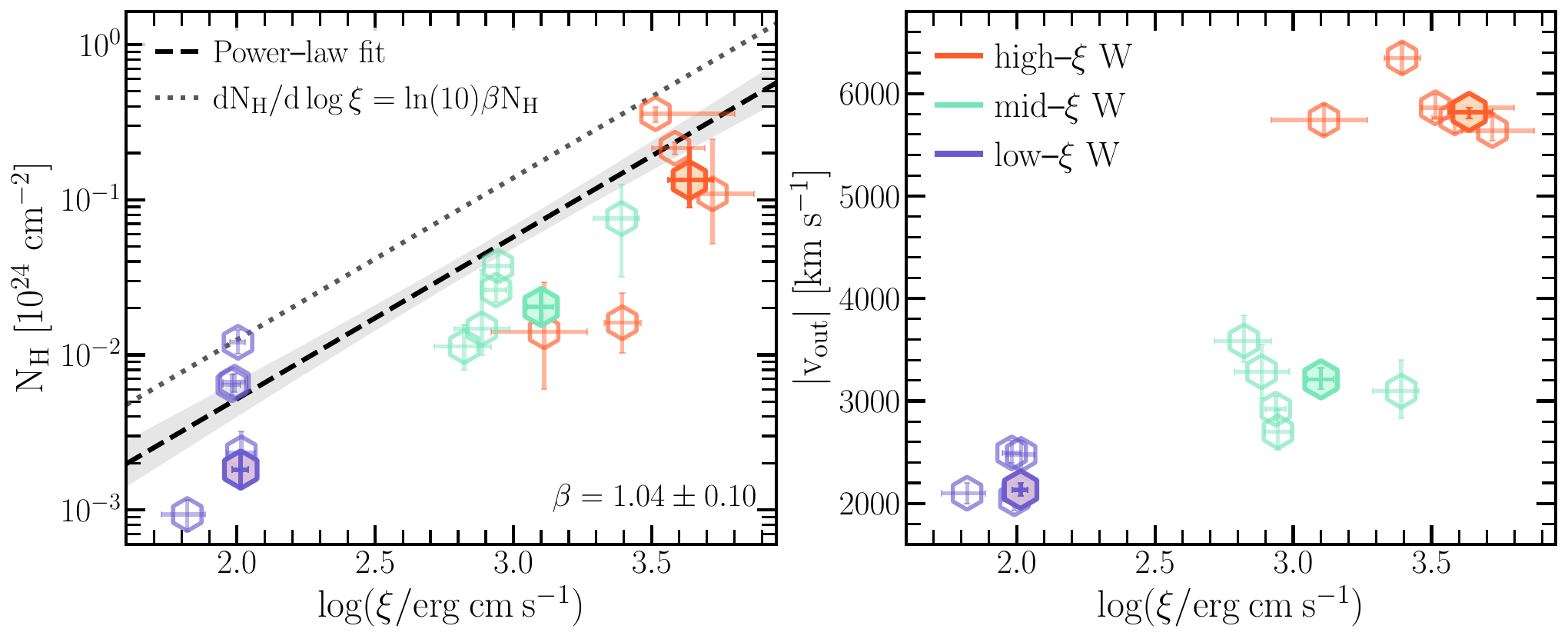}

\caption{Outflow parameter correlations for Mrk~335. 
Left: column density $N_{\rm H}$ versus ionisation parameter $\log\xi$. The dashed black curve shows the best-fitting power-law relation, $N_{\rm H}\propto\xi^\beta$, fitted only to the individual-epoch measurements, with the shaded region showing the $1\sigma$ uncertainty. The dotted curve shows the corresponding smoothed absorption measure distribution, ${\rm AMD}={\rm d}N_{\rm H}/{\rm d}\log\xi$. 
Right: absolute outflow velocity $|v_{\rm out}|$ versus $\log\xi$. Colours denote the three wind phases: high-, mid-, and low-$\xi$. Open hexagons show the best-fit values for the individual epochs, while filled hexagons show the best-fit values for the stacked deep RGS spectrum. The scatter of the open symbols around the filled points provides a compact visual summary of epoch-to-epoch variability superposed on an underlying ionisation and velocity stratification.}

\label{fig:amd}
\end{figure*}

Figure~\ref{fig:amd} shows the location of the three wind components in the $(\log\xi, N_{\rm H})$ and $(\log\xi, |v_{\rm out}|)$ planes. The filled hexagons represent the best-fit parameters from the combined spectrum, while the open symbols show the values measured in the individual epochs, with their uncertainties. The three components occupy distinct regions of parameter space and show a clear stratification: both $N_{\rm H}$ and $|v_{\rm out}|$ increase with ionisation.

The left panel can be interpreted as a discrete sampling of the absorption measure distribution (AMD; \citealt{Holczer07,Behar09}), defined as ${\rm AMD}\equiv {\rm d}N_{\rm H}/{\rm d}\log\xi$. To obtain a smoothed estimate, we fitted the individual-epoch measurements with a power-law relation, $\log N_{\rm H}=A+\beta\log\xi$, or equivalently $N_{\rm H}\propto\xi^\beta$, and derived the AMD analytically as ${\rm AMD}={\rm d}N_{\rm H}/{\rm d}\log\xi=\ln(10)\,\beta\,N_{\rm H}$. This provides a model-dependent, smoothed representation of the column-density distribution rather than a non-parametric AMD reconstruction.

The best-fitting slope, $\beta=1.04\pm0.10$, indicates that the column density per decade in ionisation increases toward higher $\xi$, implying that the absorber is weighted toward the more ionised phases. This is steeper than the relatively flat AMDs inferred for classical Seyfert warm absorbers, where typical slopes are $0<a<0.4$ \citet{Behar09, Laha14, Adhikari15,Keshet22}. This difference suggests that the obscurer in \mrk\ is dominated by higher-ionisation, higher-column, and faster gas, pointing to a more compact and dynamically active part of the disk wind rather than to a standard extended warm absorber. A qualitatively similar picture is emerging from recent \xrism\ observations of powerful UFOs such as PG~1211+143 and PDS~456, where the outflows are resolved into multiple ionisation and velocity components, consistent with clumpy and stratified disk-wind structures \citep{Xrism25a,Xu25,Reeves26b}. Although the \mrk\ obscurer is slower and less extreme, the steeply rising AMD and the increase of $|v_{\rm out}|$ with $\xi$ suggest that it may trace a dynamically active inner-wind region rather than a classical extended warm absorber.

Under the simplifying assumption that the clumpy phases trace an underlying radial stratification, the AMD can be used to estimate an effective density profile. This is similar to the hybrid picture proposed for PG~1211+143, in which a large-scale wind is launched over a range of radii while denser clumps produce part of the observed ionisation structure \citep{Reeves26b}. Following \citet{Behar09}, if ${\rm AMD}\propto\xi^\beta$ and the flow is approximated by an effective radial density profile $n(r)\propto r^{-\alpha}$, then $\alpha=(1+2\beta)/(1+\beta)$. In our case, the fitted slope gives $\alpha\simeq1.5$, corresponding to $n(r)\propto r^{-1.5}$. This value is close to the $\alpha=1.5$ scaling expected for Blandford--Payne-like self-similar magnetocentrifugal winds \citep{Blandford82,Contopoulos94}. The rising AMD, together with the increase of $|v_{\rm out}|$ with $\xi$, is therefore qualitatively aligned with a magnetically driven disk-wind interpretation \citep[e.g.][]{Fukumura10,Fukumura15,Fukumura21,Mehdipour25}, although it does not uniquely determine the launching mechanism.

%-----------------------------------------------------------------
%											NH vs XI and v vs Xi
%-----------------------------------------------------------------

%-----------------------------------------------------------------
%                                                    RGS spectrum
%-----------------------------------------------------------------
\begin{figure}[t!]
\centering
\includegraphics[width=\hsize]{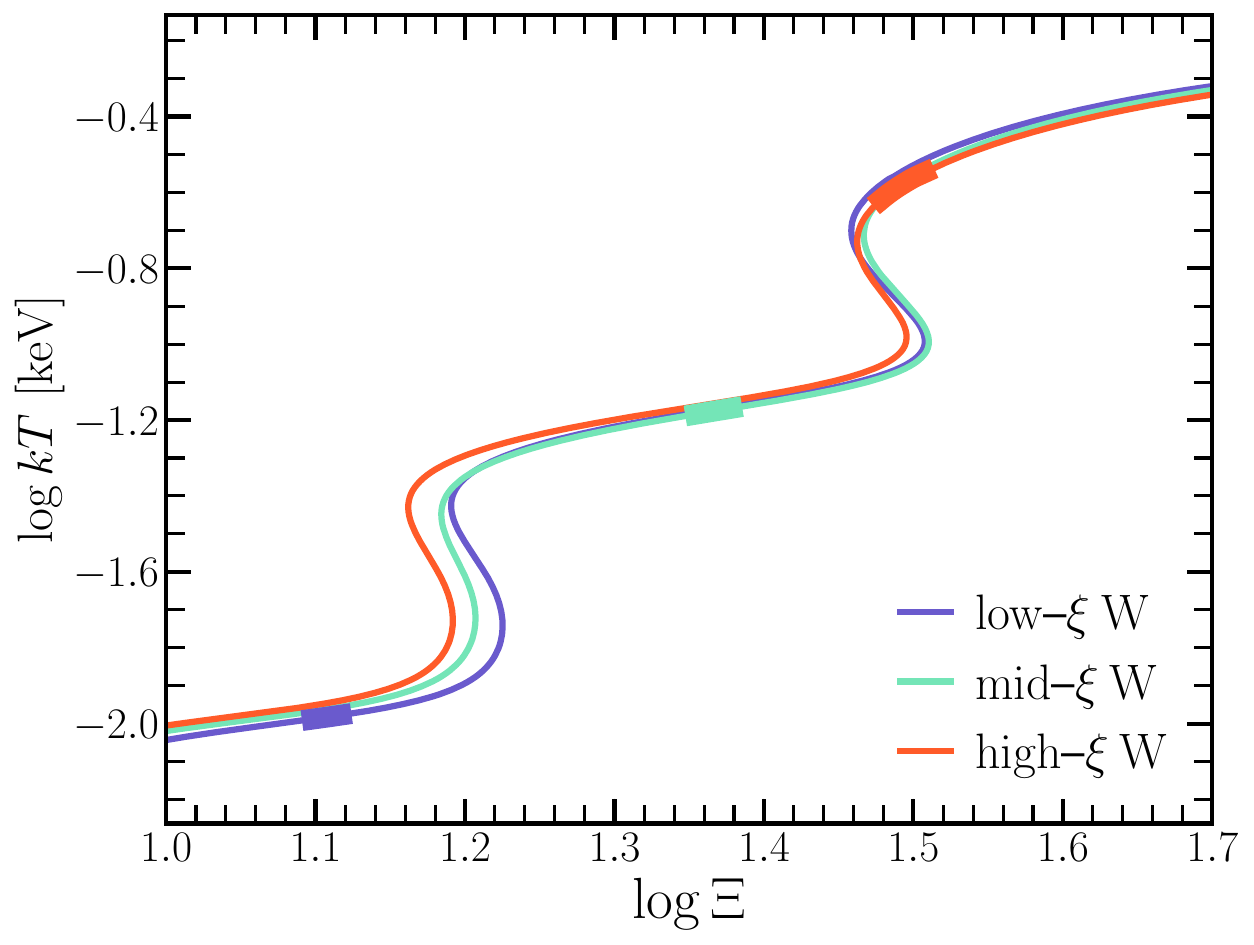}
\caption{Thermal stability curves (S-curves) for each ionised absorber, computed for the SED seen by each component, and the location of the absorbers assuming thermal equilibrium. Marker sizes reflect the $1\sigma$ uncertainty on $\log\xi$.}
\label{fig:scurve}
\end{figure}
\subsubsection{Thermal stability}
\label{sec:thermal_stability}

The thermal stability of the absorbers can be assessed using the stability curves shown in Fig.~\ref{fig:scurve}. Because the three components are arranged along the line of sight, each absorber sees a different ionising SED owing to shielding by the components closer to the central source; we therefore compute a separate stability curve (S-curve) for each absorber using the corresponding transmitted SED. These curves summarise the thermal equilibrium solutions of photoionised gas: for a given radiation field they indicate which combinations of temperature and ionisation/pressure can be maintained in steady state, and whether small perturbations in temperature will damp out (stable) or run away (unstable). The curves are expressed in terms of the pressure ionisation parameter $\Xi$, defined as the ratio of radiation pressure to gas pressure, $\Xi = \xi/(4\pi c kT)$ \citep{Krolik81}, where $c$ is the speed of light and $kT$ is the electron temperature. Regions with negative slope on the S-curve correspond to thermally unstable equilibrium, whereas positive slopes indicate thermally stable solutions. 

Given the inferred persistence of the obscurer (Sections~\ref{sec:short_term} and~\ref{sec:longterm}), thermal stability is expected. The locations of the three components are consistent with stable branches of their respective curves. Notably, the high-$\xi$ component lies close to the turning point where the curve transitions between unstable and stable branches. Gas near this critical point can undergo large temperature/ionisation changes in response to modest variations in the incident ionising flux or shielding, naturally promoting a multi-phase structure in which unstable gas fragments into cooler clumps embedded in a hotter, more tenuous medium \citep{Dannen20}. This offers a natural interpretation for the rapid evolution of the high-$\xi$ phase during the campaign: in Epoch~2, the best-fit ionisation parameter decreases (Table~\ref{tab:outflows}), consistent with the high-$\xi$ gas moving off the turning point toward a cooler, less ionised equilibrium branch as the incident ionising continuum and/or shielding conditions change.

\subsubsection{Comparison with other obscurers}
\label{sec:compare_obscurers}

A remarkably similar structure of the obscurer studied here was reported in Mrk~817 during the AGN~STORM~2 campaign \citep{Zaidouni24}, with three ionisation components ($\log\xi \approx 2.31$, $2.90$, and $3.46$), outflow velocities of $\sim$3900--5900~$\mathrm{km~s^{-1}}$, and column densities of order $10^{21}$--$10^{22}~\mathrm{cm^{-2}}$. The close resemblance in ionisation structure, column densities, and velocity stratification between \mrk\ and Mrk~817 suggests a common physical origin for these obscurers \citep{Zaidouni24}.

By contrast, \objectname{NGC\,3227} shows evidence for a two-phase obscurer during an obscured 2016 observation \citep{Wang22}, with $\log\xi \sim 1.0$ and $2.8$. In that case, the stronger absorption hampers sensitivity to higher-ionisation phases and the outflow velocities are not well constrained. Unlike \mrk\ and Mrk~817, \objectname{NGC\,3227} also hosts multiple persistent, slower warm-absorber components (typical velocities of a few hundred to $\sim$1000~km~s$^{-1}$). The apparent absence of classical, long-lived warm absorbers in \mrk\ and Mrk~817, even during high-flux, nearly unabsorbed states, may indicate that obscuration in these sources is dominated by intermittent BLR-scale disk-wind episodes, rather than occurring on top of a persistent warm-absorber system as in sources such as NGC~5548 and NGC\,3227 \citep{Mehdipour22,Wang22}. It will be important to establish whether this difference is driven primarily by geometry (e.g.\ line-of-sight inclination) or by intrinsic properties of the accreting system (e.g.\ Eddington ratio, black-hole mass, or SED shape). A follow-up study will explore the population of obscurers in Seyfert~1 galaxies and search for such trends.

Interestingly, the outflow velocities of these three obscurer components are comparable to the moderately fast, highly ionised winds recently detected in the Fe\,K band of several Seyfert~1 galaxies (e.g. NGC~4151, MCG--6-30-15) with \xrism \citep[e.g.][]{Xiang25,Brenneman25}. This similarity hints at a possible connection between the soft X-ray obscurer and hotter phases traced by Fe\,K absorption, suggesting a continuous outflow spanning a broad range of ionisation. A \xrism \citep{Tashiro25} observation of \mrk\ would provide a decisive test by resolving the Fe~K band and determining whether the same velocity-stratified outflow extends into more highly ionised gas, or whether additional faster components such as UFOs are present.

\subsubsection{Testing collisional ionisation}
\label{sec:cie_test}

Finally, we tested whether the absorption could be produced by collisionally ionised gas rather than by photoionisation. In some situations, collisionally ionised and photoionised plasmas can produce qualitatively similar sets of absorption lines, especially when only a limited set of ions is detected. Collisionally ionised gas may arise from shocks generated when a fast outflow interacts with slower material, such as previously ejected components or circumnuclear/circumgalactic gas. Recent studies have reported evidence for such collisionally ionised phases in Seyfert~1 galaxies, including NGC~4051, Mrk 1239, and Mrk~766 \citep{Ogorzalek22,Buhariwalla23,MatamoroZatarain25}.

To assess this possibility, we replaced the \pion components with \texttt{hot} components in \spex\ (collisional ionisation equilibrium) for both the stacked RGS spectrum and the individual-epoch fits, keeping the overall continuum description unchanged. For each \texttt{hot} component, we fitted the column density, temperature, covering fraction, and turbulent and outflow velocities. In all cases, the fits were significantly worse than the corresponding photoionisation models, failing in particular to reproduce the relative strengths of lines from different charge states even when multiple temperature components were allowed. Quantitatively, the best-fit C-statistic increased by $\Delta C \simeq 40$--50 for each epoch when using collisional models in place of \pion, for the same number of free parameters.

We also explored hybrid models that include both photoionised (\pion) and collisionally ionised (\texttt{hot}) absorbers. These fits consistently preferred the \pion components, with the collisional components either unconstrained or driven to negligible column densities. The hybrid fits yield only marginal improvements ($\Delta C \simeq 0$--10) despite introducing additional free parameters (six in total). We therefore conclude that the obscurer in \mrk\ is statistically (and physically) better described as photoionised gas, as expected in the presence of the strong AGN radiation field. This interpretation is also consistent with previous analyses of \mrk\ in obscured/intermediate states \citep[e.g.][]{Longinotti13} and with the long-term behaviour discussed in Section~\ref{sec:longterm}.

\subsection{Origin and energetics} 
Classical warm absorbers in Seyfert~1 galaxies are typically inferred to arise on scales of $\sim0.1$~pc to $\sim1$~kpc and have characteristic outflow velocities of a few hundred to $\sim10^{3}$~km~s$^{-1}$. In \mrk, the three photoionised absorbers detected in the intermediate-flux state have substantially higher velocities ($|v_{\rm out}|\simeq2100$--$5800$~km~s$^{-1}$; Table~\ref{tab:outflows}), pointing to an origin closer to the accretion flow/BLR region and disfavoring a torus-scale location. The similarity in ionisation structure and kinematics with the obscurer in Mrk~817 \citep{Zaidouni24} further suggests that clumpy, multi-phase obscurers may represent a common phenomenon in Seyfert~1 galaxies. To enable a direct comparison with Mrk~817, we follow the same order-of-magnitude arguments of \citet{Zaidouni24} to constrain the location (and, below, the energetics) of the outflow.

\subsubsection{Distance}
\label{sec:origin_distance}

\paragraph{Upper limit from thickness arguments.}
An upper limit on the distance can be obtained by requiring that the absorber thickness ($\Delta r$) does not exceed its distance from the source, $\Delta r/r<1$. Using $N_{\rm H}\sim n\,C_{\rm v}\,\Delta r$ and substituting $n=L_{\rm ion}/(\xi r^{2})$ gives
\begin{equation}
r_{\rm max} \simeq \frac{L_{\rm ion}}{N_{\rm H}\,\xi},
\end{equation}
where $C_{\rm v}<1$ is a volume-filling factor. Using the average ionising luminosity from the SED modelling (Table~\ref{tab:sed}, $L_{\rm ion}\simeq1.35\times10^{44}$~erg~s$^{-1}$) yields $r_{\rm max}\sim0.09$~pc for the high-$\xi$ component and $r_{\rm max}\sim2$~pc for the mid-$\xi$ component. For the low-$\xi$ component, this constraint is comparatively weak  ($\sim 300$~pc) because of its smaller $\xi$ and $N_{\rm H}$. These limits are conservative, since $C_{\rm v}$ is expected to be well below unity in a clumpy wind. In addition, the differences in the transmitted ionising luminosity seen by the individual components are modest ($\lesssim$ a few per cent), so adopting the average $L_{\rm ion}$ is sufficient for these order-of-magnitude estimates.

A characteristic radius can be estimated by assuming that the measured outflow velocity is comparable to the local escape velocity, $v_{\rm out}\approx v_{\rm esc}=\sqrt{2GM_{\rm BH}/r}$. This provides only an order-of-magnitude reference: the wind need not move at the escape speed, since it may be a failed wind (with a total velocity below $v_{\rm esc}$) or, conversely, a successful wind whose total velocity exceeds $v_{\rm esc}$, while the observed line-of-sight component depends on geometry. Under this assumption,
\begin{equation}
\frac{r}{R_{\rm g}} \simeq 2\left(\frac{c}{|v_{\rm out}|}\right)^{2},
\end{equation}
where $R_{\rm g}=GM_{\rm BH}/c^{2}$. Adopting $M_{\rm BH}=2.7\times10^{7}\ M_{\odot}$ \citep{Grier12}, we obtain
$r\simeq5.2\times10^{3}\,R_{\rm g}$ ($\simeq6.7\times10^{-3}$~pc) for the high-$\xi$ component ($|v_{\rm out}|\simeq5800$~km~s$^{-1}$),
$r\simeq1.8\times10^{4}\,R_{\rm g}$ ($\simeq2.3\times10^{-2}$~pc) for the mid-$\xi$ component ($|v_{\rm out}|\simeq3200$~km~s$^{-1}$),
and $r\simeq3.9\times10^{4}\,R_{\rm g}$ ($\simeq5.0\times10^{-2}$~pc) for the low-$\xi$ component ($|v_{\rm out}|\simeq2100$~km~s$^{-1}$).
These radii lie naturally on BLR scales. 

Using the kinematic estimate for the low-$\xi$ component, the implied number density is $n \sim 5\times10^{7}\ \rm cm^{-3}$. At this density, the recombination timescales of the dominant \xmm/RGS\ ions (e.g. \ovii--\oviii, \neix--\nex, and the Fe UTA/L-shell ions) are short, of order hours. A $\sim 1$~day response would instead require a lower density, $n \sim 5\times10^{6}\ \rm cm^{-3}$, implying a larger distance ($r \sim 0.1$--$0.2$~pc) for the low-$\xi$ phase. If the tentative $\sim$1-day delay of the low-$\xi$ component is confirmed (see Section~\ref{sec:short_term}), the absorber would lie beyond the escape-radius estimate, disfavouring a failed-wind interpretation.

\citet{Zaidouni24} used short-timescale variability and an assumed compact corona of size $R_{\rm c}\sim10\,R_{\rm g}$ to estimate a transverse velocity
$v_{\perp}\gtrsim R_{\rm c}/\delta t$, and (assuming Keplerian motion) infer $r/R_{\rm g}\sim(c/v_{\perp})^{2}$.
Here $\delta t$ denotes the characteristic timescale over which a significant change in the absorber is observed (e.g.\ in $N_{\rm H}$ and/or $\xi$), interpreted as the time required for a clump to traverse a distance comparable to the projected size of the X-ray corona. In \mrk, the strongest absorber changes occur on day timescales across the five epochs (Section~\ref{sec:short_term}). Note that the clump does not have to cross the entire X-ray source size, just a significant portion of it.
For $\delta t\sim1$~day and $R_{\rm c}\sim10\,R_{\rm g}$, the implied transverse velocity is $v_{\perp}\sim4600$~km~s$^{-1}$, corresponding to
\begin{equation}
r \sim 4\times10^{3}\,R_{\rm g} \simeq 5\times10^{-3}\ \mathrm{pc}.
\end{equation}
This estimate is consistent with the escape-velocity radii above and points to a location comparable to the characteristic H$\beta$ BLR radius of \mrk,
$R_{\rm BLR}\simeq14$ light-days $\approx 10^{4}\,R_{\rm g}$ \citep{Grier12}. Given the sparse sampling, it should be regarded as indicative.

Taken together, these arguments place the obscurer within $\sim10^{-3}$--$10^{-1}$~pc ($\sim10^{3}$--$10^{5}\,R_{\rm g}$), with the high-$\xi$ phase located closest to the black hole and the low-$\xi$ component further out. This ordering is consistent with the short-term variability trends (Section~\ref{sec:short_term}), where the high-$\xi$ component responds most directly to changes in $L_{\rm ion}$, while the low-$\xi$ phase shows at most a delayed response. Our inferred distance range is broadly consistent with the interpretation of the 2009 intermediate-flux state in \citet{Longinotti13}, who also reported kinematic consistency between the UV and X-ray absorbers on BLR-like scales. The distances inferred here are also comparable to those obtained for Mrk~817 by \citet{Zaidouni24}, which is particularly notable given the similar black-hole masses of the two systems (for Mrk~817, $\log(M_{\rm BH}/M_{\odot})=7.59$; \citealt{Bentz18}).

\subsubsection{Energetics}
\label{sec:origin_energetics}
Following \citet{Zaidouni24}, we estimate the mass outflow rate and kinetic power of the obscurer with simple continuity arguments. For a radial outflow subtending a solid angle $\Omega$, the mass flux is $\dot{M}=\Omega r^{2}\rho v$, where $v$ is the outflow velocity and $\rho=C_{\rm v}m_{\rm p}n$ is the mass density (with particle density $n$, proton mass $m_{\rm p}$, and volume-filling factor $C_{\rm v}<1$). The observed column density is $N_{\rm H}\simeq n\,C_{\rm v}\Delta r$, where $\Delta r$ is the absorber thickness.

Requiring that the absorber is not geometrically thicker than its distance from the source ($\Delta r<r$) gives a lower limit on the mass outflow rate per unit solid angle that is independent of $C_{\rm v}$,
\begin{equation} 
\begin{aligned} \left(\frac{\dot{M}_{\rm min}}{\Omega}\right)_{\rm tot} &\gtrsim m_{\rm p}\sum_i r_i N_{{\rm H},i}|v_i| \\ 
&\simeq 5.6\times10^{-2}\, M_{\odot}\,\mathrm{yr^{-1}\,sr^{-1}}, 
\end{aligned} \label{eq:mdotmin_sum} 
\end{equation}
where the sum runs over the three \pion components, using their time-averaged $N_{\rm H}$ and $v_{\rm out}$ (Table~\ref{tab:outflows}). A characteristic radius for each phase is assigned using the escape-velocity estimate (Section~\ref{sec:origin_distance}), $r_{i}/R_{\rm g}\simeq 2(c/|v_{i}|)^{2}$. This mapping is uncertain because the true 3D velocity field and geometry are not directly measured.

The corresponding minimum kinetic luminosity follows from $L_{\rm kin}=\frac{1}{2}\dot{M}v^{2}$:
\begin{equation}
\begin{aligned}
\left(\frac{L_{\rm kin,min}}{\Omega}\right)_{\rm tot}
&\gtrsim \frac{1}{2}m_{\rm p}
\sum_i r_i N_{{\rm H},i}|v_i|^3 \\
&\simeq 4.9\times10^{41}\,
\mathrm{erg\,s^{-1}\,sr^{-1}} \\
&\simeq 0.12\%\,L_{\rm bol}\,sr^{-1},
\end{aligned}
\label{eq:lkinmin_sum}
\end{equation}
where $L_{\rm bol}=4.07\times10^{44}\ \mathrm{erg\ s^{-1}}$ is calculated from our unobscured SED model integrated over $10^{-3}$--$10^{5}$~eV.

An upper-limit scaling can be obtained by eliminating $n$ using the ionisation parameter $\xi=L_{\rm ion}/(nr^{2})$ \citep{Blustin05}. This yields, for each phase, $\dot{M}/(C_{\rm v}\Omega)\simeq m_{\rm p}L_{\rm ion}|v|/\xi$, and therefore
\begin{equation}
\begin{aligned}
\left(\frac{L_{\rm kin}}{C_{\rm v}\Omega}\right)_{\rm tot}
&\simeq \frac{1}{2}m_{\rm p}L_{\rm ion}
\sum_i\frac{|v_i|^3}{\xi_i} \\
&\simeq 1.9\times10^{43}\,
\mathrm{erg\,s^{-1}\,sr^{-1}} \\
&\simeq 4.7\%\,L_{\rm bol}\,sr^{-1},
\end{aligned}
\label{eq:lkin_upper}
\end{equation}
using time-averaged $\xi_{i}$ and $|v_{i}|$ from Table~\ref{tab:outflows} and
$L_{\rm ion}=1.35\times10^{44}$~erg~s$^{-1}$ (Table~\ref{tab:sed}).

We adopt a fiducial global covering factor $\Omega/4\pi=0.25$ (i.e.\ $\Omega=\pi$). The volume-filling factor is unknown; within the upper-limit scaling, a conservative bound is obtained by taking $C_{\rm v}=1$. For these assumed values, we obtain $0.38\% \lesssim L_{\rm kin}/L_{\rm bol} \lesssim 15\%$. More realistically, if $C_{\rm v}\lesssim0.08$ as inferred for Seyfert warm absorbers \citep{Blustin05}, the scaling gives $L_{\rm kin}/L_{\rm bol}\lesssim1.2\%$. Comparing with the commonly adopted feedback threshold $L_{\rm kin}/L_{\rm bol}\sim0.5\%$ \citep{Hopkins10}, often used as an approximate criterion for an outflow to carry enough kinetic power to affect the host galaxy on large scales, the upper-limit scaling exceeds this threshold for $C_{\rm v}\gtrsim0.034$ (for $\Omega=\pi$), while the lower bound already requires $L_{\rm kin}/L_{\rm bol}\gtrsim0.38\%$. This might suggest that the obscurer in \mrk\ could be energetically significant for AGN feedback, depending on its global covering and clumpiness.

These energetics are similar to those inferred for Mrk~817 \citep{Zaidouni24}. We stress, however, that they remain uncertain because they depend on the poorly constrained global geometry ($\Omega$) and volume filling factor ($C_{\rm v}$), as well as on the simplifying assumptions used to relate velocity to radius in the lower-limit estimate. Moreover, the kinetic power inferred here is based on a limited number of epochs and may not represent the long-term duty cycle of the obscurer.

\subsection{Emission lines} \label{sec:emission_lines}
The RGS spectrum of \mrk\ shows clear signatures of multiple narrow emission lines. The strongest features are H-like and He-like transitions of C, N, and O, which likely originate in photoionised gas on scales larger than the immediate vicinity of the X-ray source (e.g.\ the narrow-line region and/or an extended outflow). These emission lines dominated the RGS spectrum during previous low-flux/obscured observations in 2015 and 2018/2019, and they are also detected in the intermediate-flux observations in 2007 and 2009. In the high-flux states in 2000 and 2006 they are only weakly detected, primarily because the strong continuum reduces the line contrast.

In this work, our primary focus is the obscurer seen in absorption. To keep the fits computationally efficient, we model the emission lines phenomenologically with Gaussian profiles, following the approach of \citet{Longinotti13}, rather than using a full photoionisation emission calculation with \pion. As a consistency check, we performed exploratory fits including photoionised emission components and found that two phases with $\log\xi \sim 1.0$ and $\sim 2.3$ are required to reproduce the line spectrum, broadly consistent with previous work. A dedicated analysis of the emission-line plasma, including self-consistent photoionisation emission modelling, is deferred to future work.

The full list of detected emission lines is given in Table~\ref{tab:emission_lines}. The Ly$\alpha$ and He$\alpha$ emission lines of carbon, nitrogen, and oxygen are relatively narrow, with ${\rm FWHM}\lesssim0.2$~\AA. The main exception is N\,\textsc{vii} Ly$\alpha$, whose fitted width is larger; this feature is likely affected by blending with the recombination-continuum \cvi structure \citep[e.g.][]{Parker19}. In addition, the \oviii\ and \cvi\ Ly$\alpha$ lines show broad residual emission underlying the narrow cores. We modelled these residuals by adding broad Gaussian components centred at the corresponding reference wavelengths. The origin of these broad components (FWHM $\sim$3--4~\AA) is uncertain. One possibility is that they trace reprocessed emission from the inner accretion flow (e.g.\ relativistically broadened disc emission), although our spectral setup does not include a self-consistent soft X-ray reflection model that could test this scenario. Alternative explanations include Compton broadening in an optically thick scattering region. 

In the individual-epoch fits, we allowed only the line normalisations to vary, while the line centroids and widths were fixed to their values from the stacked spectrum. We find that the line normalisations vary significantly between epochs. If the dominant emission originates at large radii, such variability on day timescales would be unexpected and may instead reflect changes in the underlying continuum/absorption affecting the inferred line fluxes, or a contribution from a more compact emitting/scattering region. A possible compact origin has already been suggested by \citet{Longinotti08}, who found that the \ovii\ triplet in the 2007 low-flux observation was consistent with densities typical of the BLR. This is comparable to the spatial scale inferred for the obscuring wind in the 2009 intermediate-flux observation \citep{Longinotti13} and in the 2021 observations analysed here, raising the possibility that at least part of the soft X-ray line emission is associated with the wind itself. We leave a detailed investigation of the emission-line variability and origin to future work, where the full line forest will be modelled with photoionisation emission codes.

\section{Conclusion} \label{sec:conclusion}
We presented results from a coordinated \xmm/\nustar\ campaign on \mrk\ in June 2021, complemented by long-term \swift\ monitoring. The source was observed in an intermediate-flux state, enabling a detailed high-resolution study of the ionised obscurer with the RGS and, for the first time in \mrk, a time-resolved characterisation of its variability on day timescales. Our main results are:

\begin{itemize}
\item \textit{Three-phase obscurer.} The stacked RGS spectrum reveals three photoionised absorption components with time-averaged ionisation parameters of $\log\xi \simeq 3.69$ (high-$\xi$), $2.97$ (mid-$\xi$), and $1.91$ (low-$\xi$), outflowing at $|v_{\rm out}|\simeq 5800$, $3200$, and $2100$~km~s$^{-1}$, respectively. Their ionisation structure and column densities are broadly consistent with those inferred during the intermediate-flux \xmm\ observation in 2009 \citep{Longinotti13}, suggesting that similar multi-phase obscuration persists in \mrk\ over decade timescales.

\item \textit{Day-scale variability and clumpiness.} Across the five consecutive RGS spectra, all three phases show variability in column density, ionisation parameter, and outflow velocity. Under the assumption of constant covering factor, the strong day-scale changes in opacity are interpreted primarily as changes in $N_{\rm H}$ and transmission, favouring a clumpy, inhomogeneous obscurer rather than a uniform absorbing screen.

\item \textit{Flare-associated changes.} The most prominent event is the sharp drop in column density of the low-$\xi$ absorber during the flare in Epoch~4, consistent with a rapid decrease in line-of-sight opacity. Together with the velocity increase observed in Epoch~5, this behaviour may indicate that the enhanced radiation field modifies the wind structure and contributes to the acceleration of the absorbing gas, particularly for the lowest-ionisation phase.

\item \textit{Different response times across phases.} The high-$\xi$ component shows the clearest response to changes in the ionising luminosity, consistent with a location closer to the central engine than the lower-ionisation phases. The low-$\xi$ component shows at most a delayed response of order $\sim$1 day, which, if confirmed with denser sampling, would place it further out.

\item \textit{AMD and wind structure.} The absorber shows a rising $N_{\rm H}$--$\xi$ trend and an increase of $|v_{\rm out}|$ with ionisation. The corresponding smoothed AMD indicates that the column density is weighted toward the high-ionisation phases, supporting an inhomogeneous disk wind with an underlying ionisation and velocity stratification, qualitatively consistent with expectations for magnetically driven disk-wind scenarios.

\item \textit{What drives the X-ray variability.} The intermediate-state X-ray variability is driven by both intrinsic continuum changes and variable absorption. In our empirical decomposition, changes in the baseline continuum dominate the broadband variability, while changes in the transmission of the modelled wind provide an additional, non-negligible modulation. The tendency for the low-$\xi$ opacity to decrease during high-luminosity phases suggests that radiation may play an important role in shaping the coldest phase of the obscurer.

\item \textit{Location and energetics.} Order-of-magnitude distance estimates place the obscurer at BLR scales, spanning $\sim10^{3}$--$10^{5}\,R_{\rm g}$, comparable to, and partly interior to, the characteristic BLR radius of \mrk. Depending on the global geometry and clumpiness, the inferred kinetic power can reach the percent level of $L_{\rm bol}$, implying that such obscurers may contribute to AGN feedback.
\end{itemize}

Future high-cadence, simultaneous UV--X-ray monitoring, combined with time-dependent photoionisation modelling, will be essential to test these distance estimates directly by measuring the absorber density from its response to continuum variability.

\begin{acknowledgments}
We thank the referee for the thorough review of the manuscript and for the helpful insights that have improved the quality of the paper. 
D.R. is grateful to Jelle de Plaa and Liyi Gu for their support and assistance with the SPEX software. During the project, D.R. received support from the Margaret Burbidge Fellowship supported by the Brinson Foundation as well as the NASA through the Smithsonian Astrophysical Observatory (SAO) contract SV3-73016 to MIT for Support of the Chandra X-Ray Center (CXC) and Science Instruments. P.K. is supported by NASA through the NASA Hubble Fellowship grant HST-HF2-51534.001-A awarded by the Space Telescope Science Institute, which is operated by the Association of Universities for Research in Astronomy, Incorporated, under NASA contract NAS5-26555. E.B. is supported by The Israel Science Foundation (grant No. 2617/25). This research is partly based on observations obtained with XMM-Newton, an ESA science mission
with instruments and contributions directly funded by
ESA Member States and NASA. 
\end{acknowledgments}


\vspace{5mm}
\facilities{Swift (XRT, UVOT), \xmm(RGS, EPIC-pn), \nustar}


\software{SPEX \citep{Kaastra20}, HEASoft, Procreate, Matplotlib, Jupyter-Notebook}



%% For this sample we use BibTeX plus aasjournalv7.bst to generate the

\bibliography{biblio.bib}{}
%\bibliography{biblio.bib}{}
\bibliographystyle{aasjournalv7}



\appendix

\section{Appendix information} \label{sec:appendix}

\begin{figure*}[h!]
\centering
\includegraphics[width=.9\hsize]{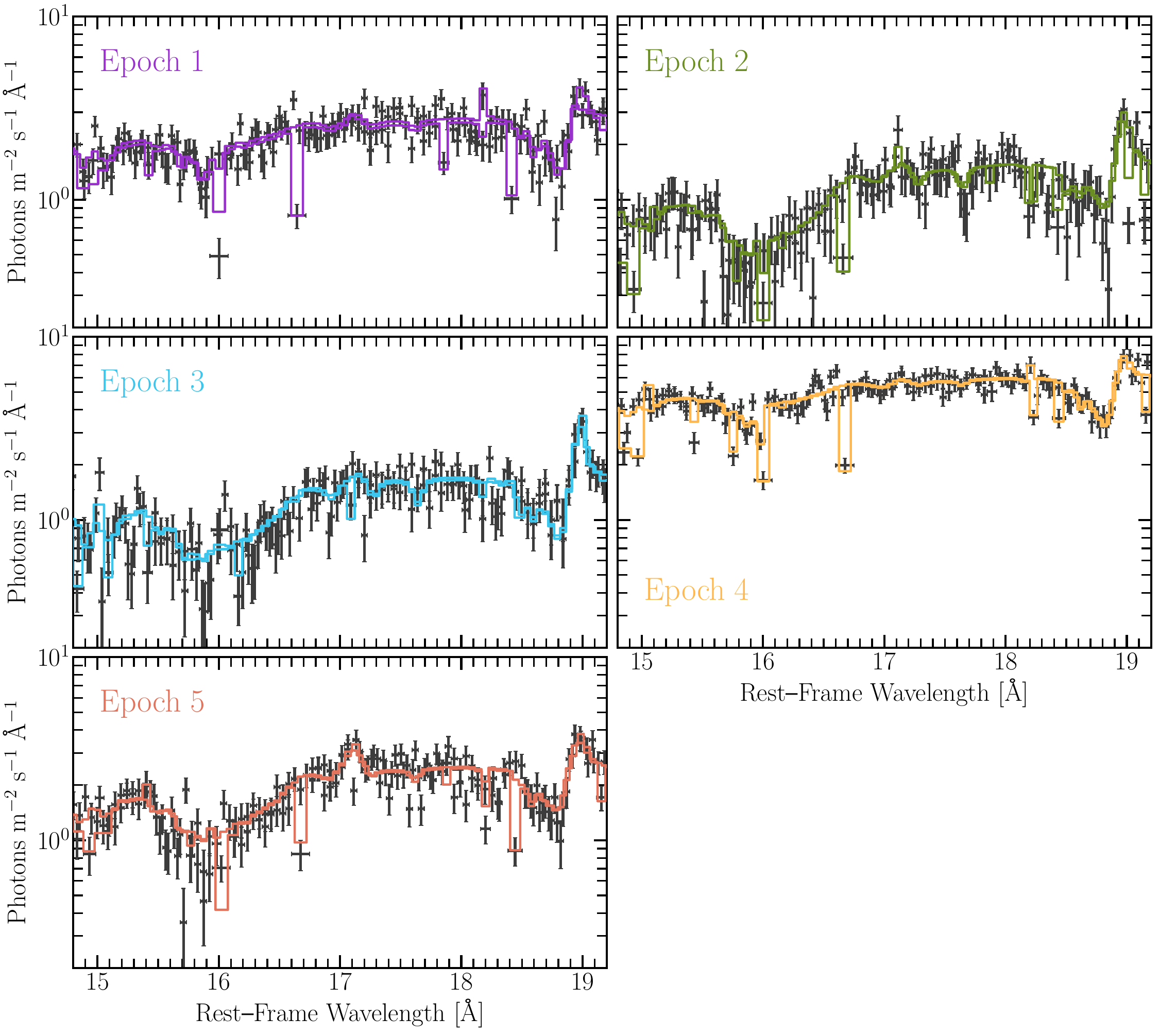}
\caption{
Time evolution of the \xmm/RGS spectra of \mrk\ in the 14.8--19.2~\AA\ rest-frame band (Fe\,\textsc{xvii}--\oviii).
Each panel shows one epoch (Epochs~1--5), with RGS1 and RGS2 fitted simultaneously; the best-fit model is overplotted.
Colors correspond to the epochs and match the palette used in the light-curve figures.
The spectra illustrate strong variability in the soft X-ray band, including changes in the continuum level and in the depth/shape of the \oviii\ Ly$\alpha$ absorption and Fe~UTA features over the $\sim$8-day campaign.}
\label{fig:rgs_evolution}
\end{figure*}

Figure~\ref{fig:rgs_evolution} provides a visual summary of the short-timescale spectral variability discussed in the main text. We display the time evolution of the RGS spectra in the Fe\,\textsc{xvii}--\oviii\ band (14.8--19.2~\AA), where the variability is most apparent. Each panel shows one epoch fitted independently (RGS1 and RGS2 fitted simultaneously), with the best-fit model overplotted. The depth of the \oviii\ Ly$\alpha$ absorption and the strength/shape of the Fe~UTA change markedly from epoch to epoch, providing a direct, model-independent view of the evolving wind transmission over the $\sim$8 days of the campaign.

%-----------------------------------------------------------------
%													RGS EPOCHS
%-----------------------------------------------------------------


\end{document}